\documentclass{mn2e}
\usepackage{graphicx,epsfig}

\newif\ifAMStwofonts

\def\deg{\hbox{$^\circ$}}

\def\fs{\hbox{$.\!\!^{\rm s}$}}

\title{XMM-Newton observations of two X-ray bright galaxy groups -- pushing
out to $r_{500}$}
\author[J. Rasmussen and T. J. Ponman]
       {J.~Rasmussen$^1$\thanks{E-mail: jr@astro.ku.dk} 
        and T.~J.~Ponman$^2$ \\
        $^1$ Astronomical Observatory, University of Copenhagen, 
        Juliane Maries Vej 30, DK-2100 Copenhagen \O, Denmark\\
        $^2$ School of Physics and Astronomy, University of Birmingham, 
        Edgbaston, Birmingham B15 2TT, UK }
\date{}

\pagerange{\pageref{firstpage}--\pageref{lastpage}}
\pubyear{2003}

\begin{document}

\maketitle

\label{firstpage}

\begin{abstract}
We report on the results of {\it XMM-Newton} observations of the 
$z=0.18$ galaxy group WARPJ0943.7+1644 and the
nearby poor cluster WARPJ0943.5+1640 at $z=0.256$. Tracing the X-ray gas out
to  $\sim 425h^{-1}$  and $\sim 500h^{-1}$ kpc in the two systems, 
corresponding to roughly 70 per cent of the estimated virial radii, 
we find the surface brightness profile of both groups to be well described by
standard $\beta$-models across the entire radial range, but with a 
significantly lower value of $\beta$ in the cooler WARPJ0943.7+1644. A
Navarro-Frenk-White (NFW) model for the gas density can also provide
a good fit to the surface brightness data, but a large scale radius is 
required, and gas-traces-mass is strongly ruled out. Both systems fall close 
to the observed $L_X-T$ relation for clusters. Gas mass fractions increase 
with radius out to the radii probed, reaching extrapolated values at the 
virial radius of 0.08 and 0.09$h^{-3/2}$, respectively. 
The latter is concordant with results obtained for more massive systems. Gas 
entropy profiles show evidence for excess entropy out to at least 
$r_{500}$, and are inconsistent with predictions of simple preheating models.
This study emphasizes the need for observing galaxy groups out to large
radii, to securely establish their global properties.
\end{abstract}


\section{Introduction}\label{sec,intro}

Groups of galaxies are of fundamental importance to our understanding of the
mass distribution in the Universe. The majority of galaxies are found in 
groups (Tully 1987), and the hot intergalactic gas in groups appears likely 
to represent the dominant component of the Universal baryon content 
(Fukugita, Hogan \& Peebles 1998). However, observation of this gas presents 
a problem.
The small size of groups, compared to rich clusters, results in lower
X-ray surface brightness, with the result that whilst cluster gas in
some rich systems has been mapped out to approximately the virial
radius $r_{200}$, defined here as the radius containing a mean
overdensity of 200 with respect to the critical density, the emission
from galaxy groups is typically lost in the X-ray background at $r \la
r_{200}/3$.

The seriousness of the problem is aggravated by the fact that surface 
brightness profiles in poor systems are usually much {\it flatter} than those
in clusters (Ponman, Cannon \& Navarro 1999; Sanderson et al.\ 2003), 
probably due to the action of non-gravitational processes such as galaxy 
winds. Adopting the popular $\beta$-model parametrization for the X-ray 
surface brightness $S$,
\begin{equation}\label{eq,beta}
S(r)=S_0[1+(r/r_c)^2]^{-3\beta +1/2},
\end{equation}
galaxy groups are generally found to have $\beta \la 0.5$ (Helsdon \&
Ponman 2000), in comparison to $\beta \approx 0.67$ in clusters 
(e.g.\ Jones \& Forman 1984; Mohr et al.\ 1995).
Hence both the luminosity and gas mass formally diverge at large radius in
groups.

The low X-ray surface brightness of galaxy groups has so far precluded
the detection of the hot gas at large radii which should dominate their 
baryon content. 
It is important to derive observational constraints on the properties
of this gas to larger radii, for a number of reasons.
One can find many examples in the literature of authors concluding that
galaxy groups have low gas mass fractions (defined as the ratio of gas mass to
total mass), on the basis of an X-ray analysis extending to only 
$r\sim 200$ kpc.
While it is known that galaxy groups contain large amounts of dark
matter, it is not yet clear whether their mass-to-light ratios and 
total baryon fractions are indistinguishable from those of clusters, since 
they have not been measured to large radii. Given the dominant contribution 
of groups to
the mass function of virialised systems, this has a substantial impact on
attempts to evaluate the observed mass and baryon content of the Universe.
It is not clear how properties measured within $\sim r_{200}/3$ 
should be extrapolated to $r_{200}$ to obtain integrated values for whole
systems and provide 'templates' for the interpretation of other poor systems.
For example, it is not obviously appropriate to fit a $\beta$-model
and use this to extrapolate the X-ray luminosity or gas density distribution.
The unphysical divergence at large radius discussed above, implies that such a
profile {\it must} turn over at some radius, and one needs to know where.
In addition, the $\beta$-profile has little theoretical justification and
better descriptions might be at hand.
For example, Navarro, Frenk \& White (1995; NFW hereafter) find from
cosmological simulations that out to $r_{200}$, profiles of both 
dark matter and gas density in clusters
are better fitted by a model of the form
\begin{equation}
\rho(r) \propto \frac{1}{(r/r_s)(1+r/r_s)^2}
\label{eq,nfw}
\end{equation}
(see also Moore et~al.\ 1998).
Here $r_s$ is a characteristic scale radius related to the mass of the system.
However, these simulations do not allow for extra physics, such as galaxy 
winds and radiative cooling, which may radically alter the gas profiles, 
especially in low-mass systems. There is no substitute for actually observing
how the gas behaves at large radii.

With the arrival of {\it XMM-Newton}, we have the opportunity for the first 
time to measure the X-ray properties of galaxy groups to large radius. 
Here we report on the results of {\it XMM-Newton} observations of the galaxy 
groups WARPJ0943.7+1644 (hereafter referred to in the text as WJ943.7) and
WARPJ0943.5+1640 (hereafter WJ943.5 in the text),
aimed at mapping and characterizing the X-ray gas out
to a considerable fraction of $r_{200}$. 
These X-ray sources are close on the sky and were targeted in a single 
{\it XMM} pointing; they were both
detected serendipitously in a 9 ks {\it ROSAT} PSPC pointing as part of the 
WARPS X-ray cluster survey (Scharf et~al.\ 1997; Jones et~al.\ 1998). 
The extended X-ray emission of the systems
was also confirmed by their subsequent detection in the CfA 160 deg$^2$ 
survey (Vikhlinin et~al.\ 1998; Mullis et~al.\ 2003) and in the RIXOS survey 
(Mason et al.\ 2000). In the optical follow-up of the former, spectroscopic 
redshifts of 0.180 and 0.256 were derived for WJ943.7 and WJ943.5, 
respectively. 

In \S \ref{sec,obs} we describe the data preparation, \S \ref{sec,analysis}
deals with the methods and results of the spatial and spectral analysis of
the groups, while \S \ref{sec,mass} describes results and \S \ref{sec,discus}
implications of the mass analysis. A summary is presented in 
\S \ref{sec,concl}.

We adopt a cosmology with $\Omega_m=0.3$ and $\Omega_{\Lambda}=0.7$,
and write the Hubble parameter as $H_0=100 h$ km s$^{-1}$ Mpc$^{-1}$.

\section{Observations and data reduction}\label{sec,obs}

WARPJ0943.7+1644 and WARPJ0943.5+1640 were observed by {\it XMM-Newton} in 
two separate 
observations for a total exposure time of 22 ks. All detectors were in Full 
Frame mode, and a medium optical blocking filter was used due to the presence
of nearby bright stars.

Data preparation was carried out using the {\sc xmmsas} v5.3 software.
Screening for periods of high background was performed by inspection of 
lightcurves extracted in the 10-12 keV (for MOS) and 12-15 keV (pn) bands. 
Periods with count rates in excess of
$0.15$ cts s$^{-1}$ for MOS1/2 and $0.22$ cts s$^{-1}$ for pn were removed.
The second of the two observations was heavily affected by periods of high 
particle background. Screening for this left 7.7, 14.5, and 14.4 ks of useful
exposure time for the pn, MOS1, and MOS2 cameras, respectively, in the first
observation, while the corresponding values for the second observation were
1.0, 4.7, and 4.6 ks. 

\subsection{Imaging}

Fig.~\ref{fig,image} shows an image in the 0.4--2.5 keV band after the 
initial data filtering.
Bright point sources were subsequently identified with {\sc xmmsas} and 
excised using an exclusion radius of 25 arcsec.
Images were background subtracted prior to analysis as follows.
For each instrument and observation, a blank sky background event file
(Lumb 2002) was
generated which had undergone identical lightcurve 
filtering to 
the data, and had been coordinate transformed to match the aspect solution
of the latter. This ensured that the same CCD regions were considered when
comparing background levels in our data and the blank sky files.
The blank sky background was then determined in the 9--12 keV band within 
an annulus of inner radius 9 arcmin and outer radius 11 arcmin, and
an analogous estimate was made for our data. 
Comparing the two background levels yielded a 
scaling factor $f$ between the high--energy background in the current 
observations and in the blank sky data. The {\em total} background of the 
latter was then scaled accordingly and subtracted from the data.

\begin{figure}
\begin{center}
\epsfxsize=8.3cm
\epsfysize=8.3cm
\epsfbox{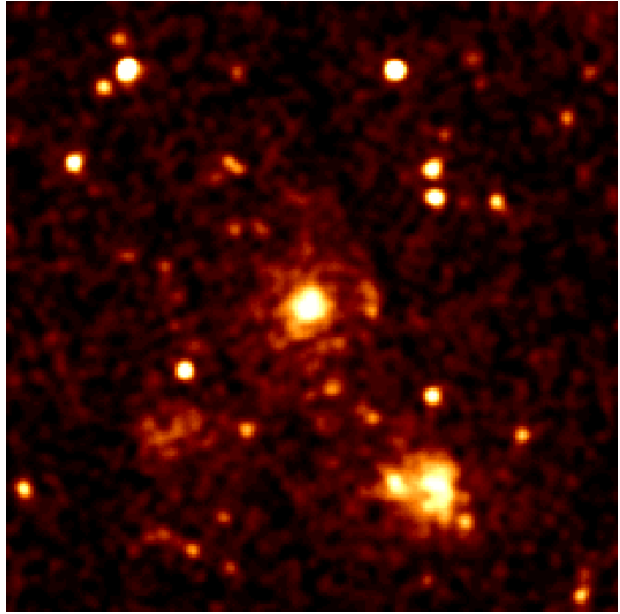}
\end{center}
\caption{Combined (pn+2MOS) 0.4--2.5 keV photon count image after initial 
data filtering. The central $15\times 15$ arcmin$^2$ is shown, smoothed 
with a Gaussian of 10 arcsec (FWHM). North is up and East is to the left.
WJ943.7 is the central extended source, while WJ943.5 is seen roughly
5 arcmin to the South-west.}
\label{fig,image}
\end{figure}

At energies $E\la 1.5$ keV, the resulting images may still contain a
contribution from the soft X-ray background, which is known to vary 
considerably with sky position and is not necessarily well represented by
the blank-sky data for a given observation.
The soft ($E<2$ keV) photon excess of the resulting images relative to the 
blank-sky background was therefore determined in a large-radius 
(9--11 arcmin) 'source-free' annulus and subtracted from the 
images after correction for vignetting. This excess was found to be negative,
i.e.\ a deficit of soft emission  relative to the blank sky files
is present in the data, constituting on average $\sim$11 per cent of the 
total background over 
the entire field of view (a deficit might be expected, as the value of 
absorbing hydrogen column density $N_H$ is on average a factor of 
$\sim$2 lower in the Lumb background data than at the position of our 
pointing).

Note that all subsequent 
analysis, including investigation of source spectra, was 
restricted to energies $E\geq 0.4$ keV, for which the effective--area 
difference is negligible between the thin optical filter used for the 
Lumb (2002) data and the medium filter applied in our observations.

\subsection{Spectra}

A similar approach to the above was used for the removal of background in
extracted spectra. Using the scaling factors $f$ derived above, source 
spectra were created, and scaled blank-sky spectra extracted from the same 
(detector) region were subtracted for each instrument and observation.
Large-radius (9--11 arcmin) source--free spectra were also 
created from the data and associated scaled blank-sky spectra were 
subtracted, the result reflecting the 
spectrum of any remaining soft background component due to the particular 
position on the sky. These residual spectra were finally also subtracted
from the source spectra. 
All results from spectral fitting presented in this paper are based on
this method for background subtraction, using the Lumb (2002) data as
background templates. However, as a check on the robustness of our results,
comparisons 
were made with results from two other 
methods for background subtraction.
One is identical to the above, but uses another set of blank-sky 
background data also assembled from pointed observations 
(Read \& Ponman 2003).
These background data have the advantage of having been obtained with the 
medium blocking filter, allowing a test for any effect which might arise
from using the thin-filter Lumb (2002) data for background subtraction in our
medium-filter observations.
The other method involves a simple local background subtraction, extracting
background spectra in a 9--11 arcmin annulus in the source data.
This ensures that the effect of any discrepancy between absorbing hydrogen
column density and optical filter in source and background data can be
tested, but it does not account for chip-to-chip variations in response or
background level. 
We find that
these two alternative methods produce results for the derived parameters 
which are consistent within $1\sigma$ with those 
obtained from the adopted approach. We therefore take our spectral results
to be fairly robust.

\section{Data analysis}\label{sec,analysis}

\subsection{X-ray/optical identification and morphology}\label{sec,morph}

As is readily apparent even in the raw imaging data (Fig.~\ref{fig,image}), 
there is evidence for a third extended extended source in the field in 
addition to the two WARPS groups: To the South-east of the optical axis,
faint extended emission is seen which was apparently not detected in 
{\it ROSAT}
pointings and is consequently not included in any {\it ROSAT}--based source 
catalogues. We shall refer to this source as XMMJ0943.9+1641.

To determine the detailed X-ray morphology of these three systems as seen by 
{\it XMM}, images were produced for each instrument and observation in the 
0.4--2.5 keV band (chosen to optimize S/N on the basis of extracted source 
spectra, see \S~\ref{sec,spectra}), from which a combined, 
background subtracted, and exposure corrected image was made. 
Fig.~\ref{fig,contours} shows resulting X-ray surface 
brightness contours of the general region surrounding WJ943.7 and of the
three extended sources in the region. Individual source contours are 
overlaid on optical images; where possible, we have used the follow-up 
imaging data from the WARPS survey itself.

\begin{figure*}
\begin{center}
\mbox{\hspace{-0.3cm}
\epsfxsize=8.3cm
\epsfysize=7.6cm
\epsfbox{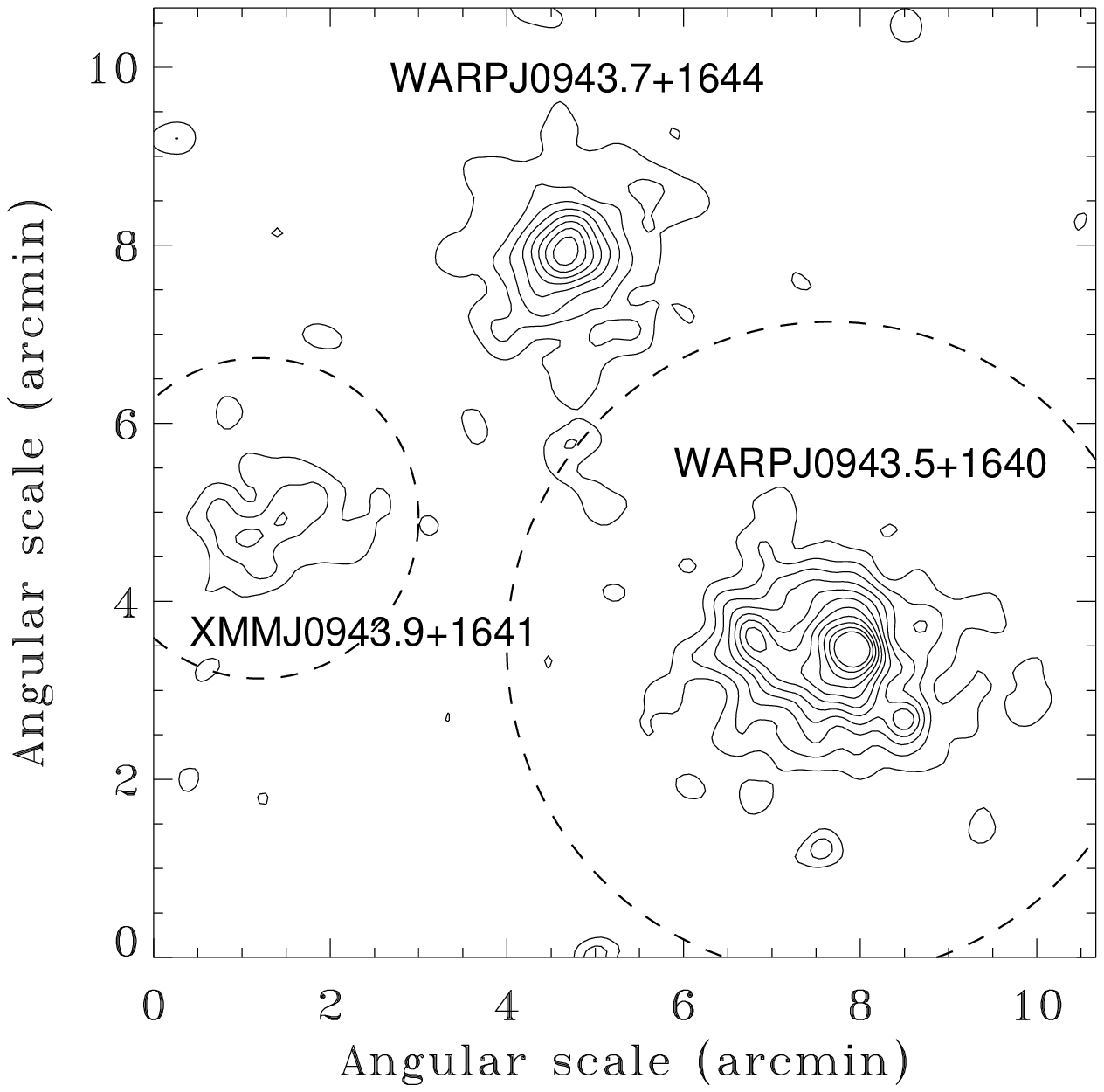}\hspace{1cm}
\epsfxsize=9.0cm
\epsfysize=7.5cm
\epsfbox{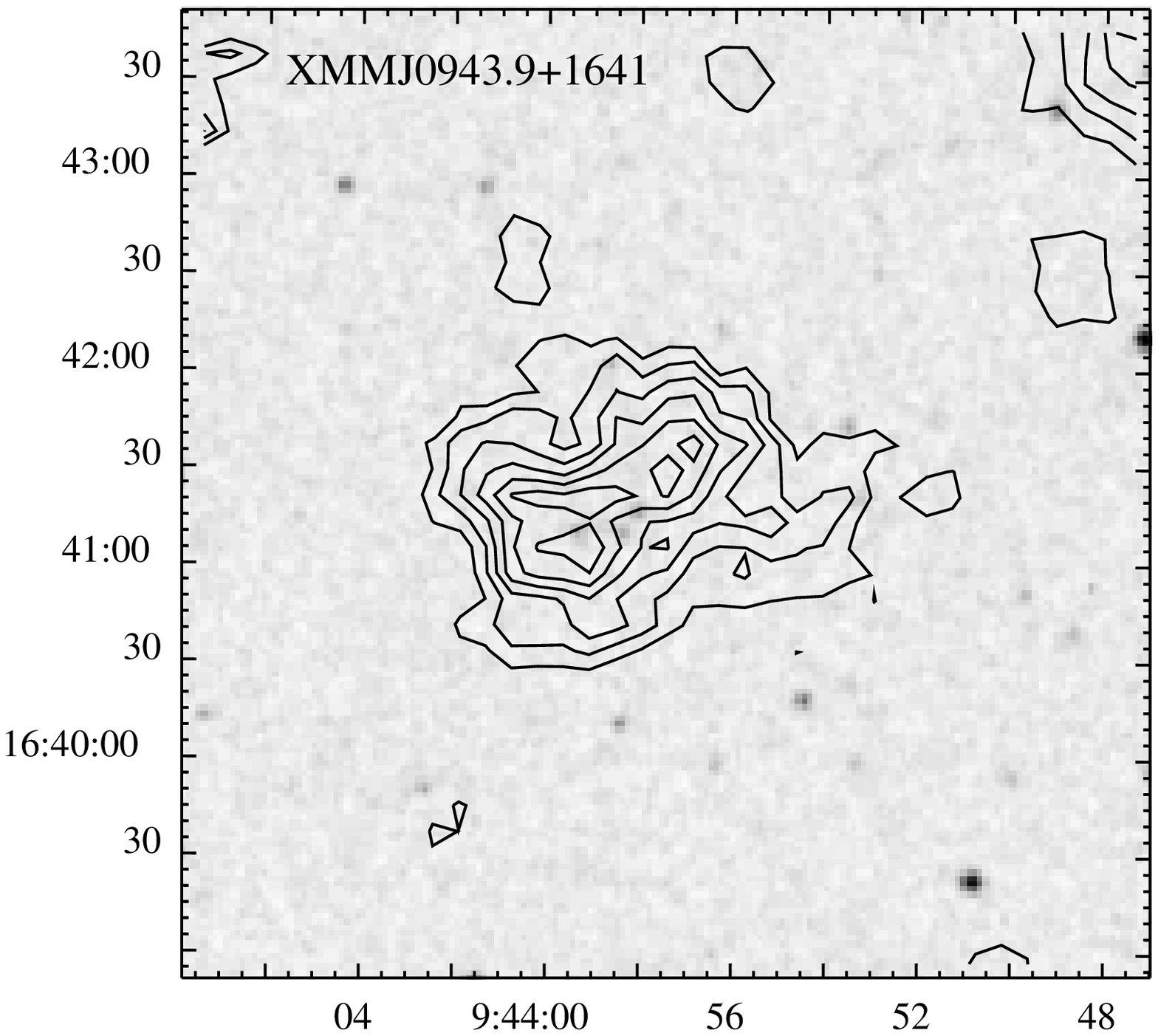}\hspace{0cm}}
\mbox{\hspace{0cm}
\epsfxsize=9.0cm
\epsfysize=7.5cm
\epsfbox{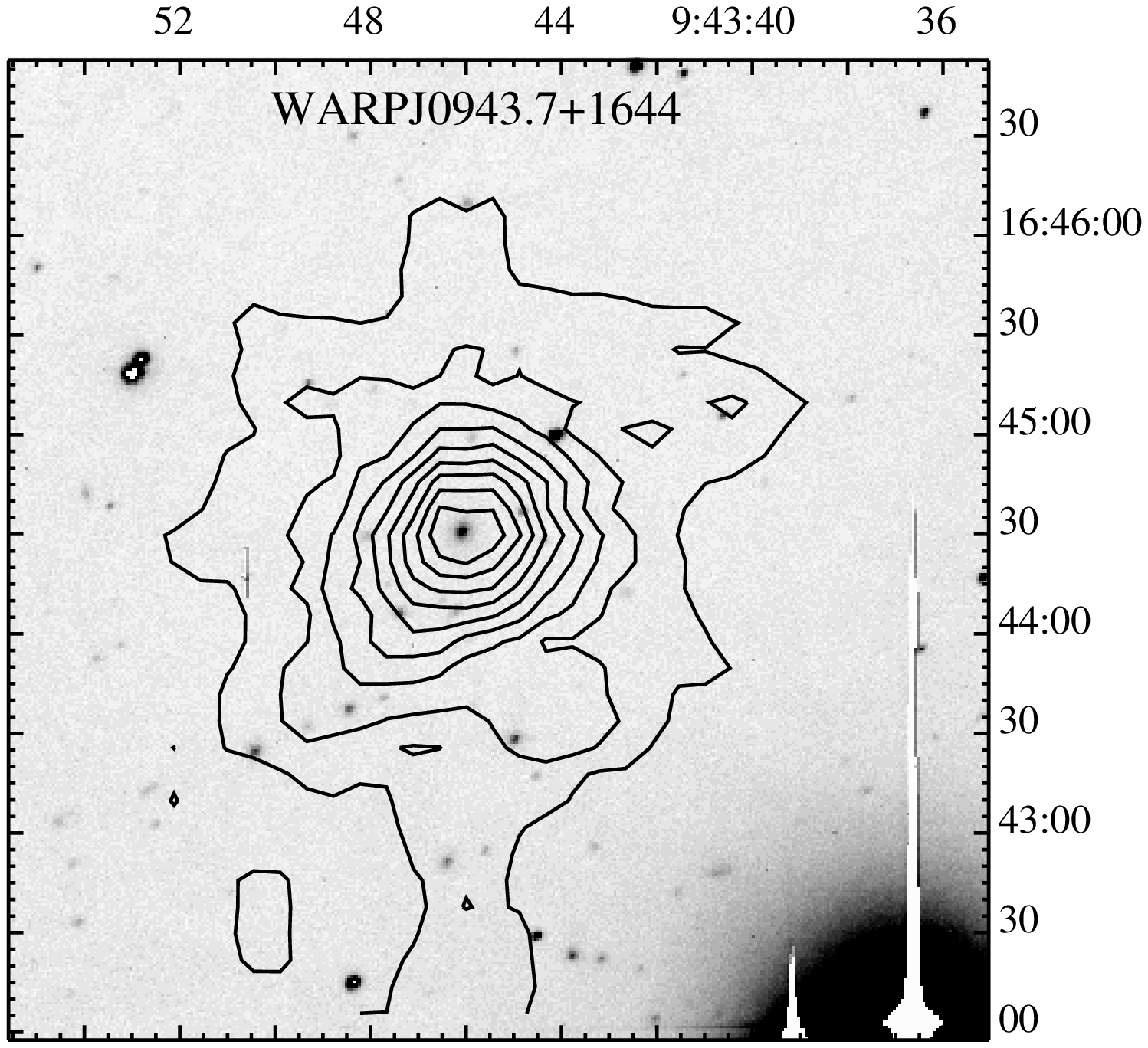}\hspace{0cm}
\epsfxsize=9.0cm
\epsfysize=7.5cm
\epsfbox{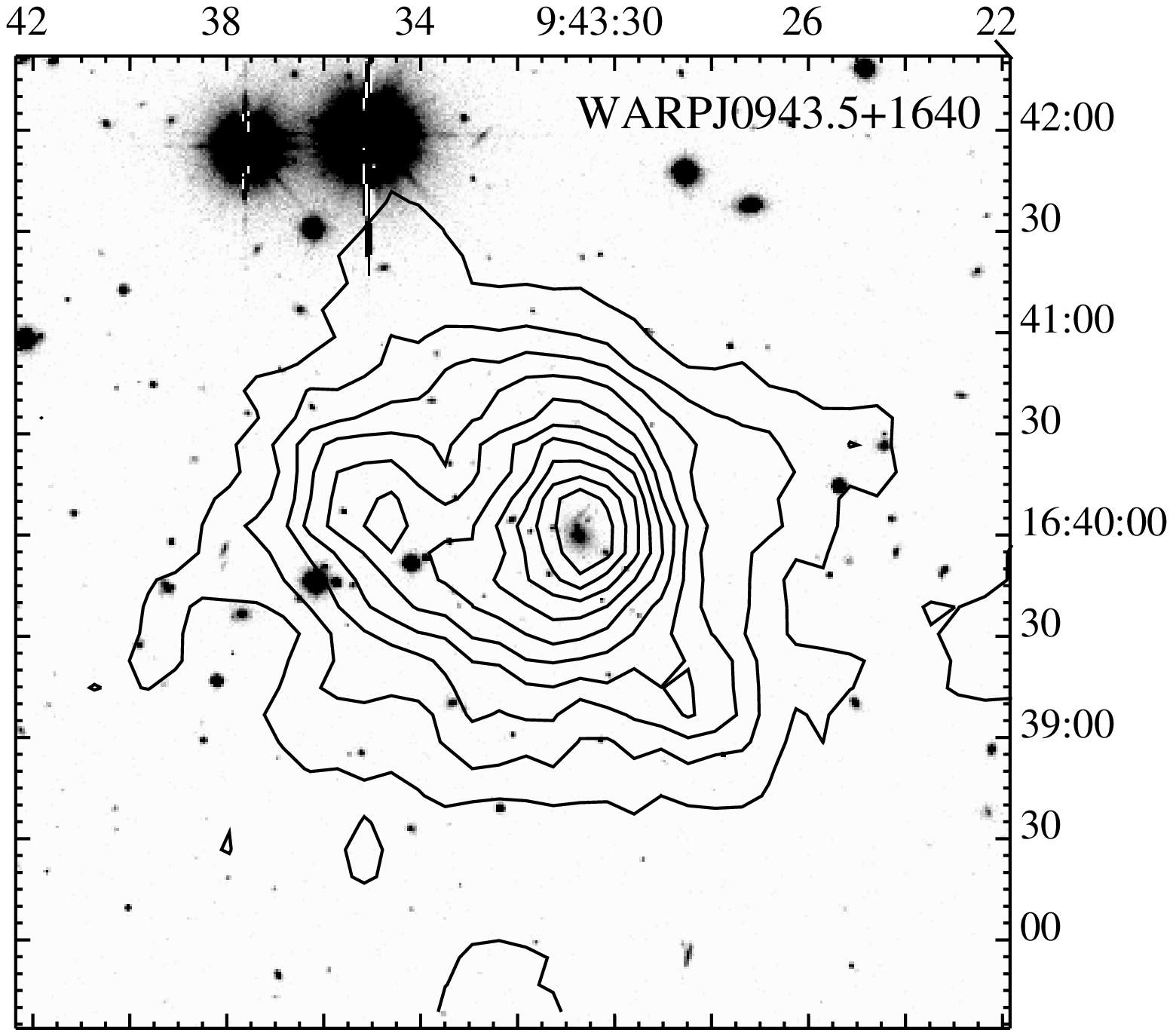}\hspace{0.0cm}}
\end{center}
\caption{{\it Top left}: 0.4--2.5 keV surface brightness contours 
of the combined (pn+2MOS) background-subtracted and
exposure-corrected image, binned to a spatial scale of 
4 arcsec pixel$^{-1}$ and smoothed with a 20 arcsec Gaussian (FWHM). 
Dashed circles mark the 
exclusion regions invoked for the surface brightness and spectral analysis
of WJ943.7 (see text). {\it Top right}:
Smoothed and linearly spaced contours 
(0.6--3 $\times 10^{-5}$ cts s$^{-1}$ pixel$^{-1}$)
of XMMJ0943.9+1641 overlayed on a
$5\times 5$ arcmin$^2$ optical image from the Digitized Sky Survey. 
{\it Bottom panel:} Contours for WJ943.7 (0.6--6 in the above units) and 
WJ943.5 (0.8--8), overlayed on WARPS survey $I$-band and $R$-band image, 
respectively, taken with the 1.3-m McGraw-Hill telescope at MDM Observatory
(images courtesy of G.\ Wegner and H.\ Ebeling).}
\label{fig,contours}
\end{figure*}

To aid in investigating the nature of optical sources in the fields, and
since spectroscopic results for the two groups are not available in the
literature, the online near-infrared ({\it JHK} bands) catalogue of the 
2 Micron All Sky Survey (2MASS) and its listing of 
associated optical sources in the Sloan Digital Sky Survey was employed. 
>From the aperture magnitudes of the sources, colour-magnitude diagrams
for all sources within $0.7h^{-1}$ Mpc of the X-ray centroid were 
produced (approximately the virial radii of the systems, see 
\S \ref{sec,mass},
corresponding to angular scales of 5.5 arcmin at $z=0.18$ and 
4.1 arcmin at $z=0.256$. 
For XMMJ0943.9+1641 a radius of 4.0 arcmin was conservatively used). 
Colours were compared to those expected for a population of passively 
evolving early-type cluster galaxies of various metallicities
at the appropriate redshift (Kodama \& Arimoto 1997).
A formation redshift of $z_f=3$ was assumed, but results are 
essentially insensitive to realistic values of this parameter.
It is worth keeping in mind when distinguishing between pointlike and 
extended NIR sources in the 2MASS catalogue that 
the point spread function (PSF) of the 2MASS telescopes is 
2.5 arcsec at FWHM, corresponding to $\sim$$5h^{-1}$ and $\sim$$7h^{-1}$ kpc
at the redshifts of WJ943.7 and WJ943.5, respectively.
At faint NIR fluxes, a source classified as pointlike could thus potentially 
be a galaxy, if only the central bulge is detectable.


\subsubsection{WJ943.7}

WJ943.7 was selected for this study in part on the basis of its 
seemingly relaxed morphology in {\it ROSAT} PSPC observations.
The overall structure of the X-ray emission is indeed seen from 
Fig.~\ref{fig,contours} to be fairly symmetric, supporting the assumption 
that the system is reasonably relaxed. 

The position of the X-ray peak and centroid at ($\alpha=09^h43^m44\fs 6, \delta=+16\deg 44'25''$) 
coincides with that of an optical source, 
clearly identified as a galaxy upon inspection of the image. No 
optical identification of this galaxy could be secured from the LEDA or NED
online databases. Its $R$-band apparent magnitude in the 2MASS catalogue is
$m_R=16.0$, making it the brightest galaxy in the field shown in 
Fig.~\ref{fig,contours}.
Assuming an early-type galaxy spectrum and 
accounting for $k-$correction (Poggianti 1997) and 
Galactic extinction (Schlegel, Finkbeiner \& Davis 1998),
this translates into 
$M_V\approx -22.9$ ($h=0.7$), which is indeed consistent with expectations
for a bright elliptical. The source displays extended 
(7 arcsec) NIR emission in the 2MASS data and we find that its 
{\it B-K}, {\it J-K}, and {\it H-K} 
colours are perfectly consistent with an elliptical galaxy at the 
spectroscopic redshift. A total of four additional extended NIR sources are 
found within $0.7h^{-1}$ Mpc, of which two have colours very similar to 
those of the central source. These four sources span an {\it R}-band apparent
magnitude range of $m_R=16.8-18.1$.
The bright source to the North-west of the X-ray centroid is a star.

\subsubsection{WJ943.5}

This source was similarly detected in both the CfA 160 deg$^2$ (Vikhlinin 
et al.\ 1998) and RIXOS (Mason et al.\ 2000) surveys.
In contrast to WJ943.7, the X-ray emission of WJ943.5 
clearly seems elongated in the NE--SW direction. The question is whether
this apparent elongation is partly or entirely due to the presence of two
nearby point sources (most clearly seen in the top
left of Fig.~\ref{fig,contours}), in particular the one East of the emission
peak. We will return to this issue in \S \ref{sec,surfbright} and 
\S \ref{sec,spectra}.
Nevertheless, the overall structure of emission appears slightly less relaxed
than for WJ943.7.

Again, the position of the X-ray peak at ($\alpha =09^h43^m31\fs 0, \delta=+16\deg 39' 57''$)
(displaced $\sim$ 15 arcsec to the W with respect to the X-ray centroid) 
coincides with an optical source. The deep optical exposure shown
in Fig.~\ref{fig,contours} allows us to readily identify it as an early-type 
galaxy. 
It is not listed in LEDA or NED but features an extended counterpart in the 
2MASS data, this being however the only 
extended NIR source within $0.7h^{-1}$ Mpc. 
Its colours are in good agreement with those expected for a cluster 
elliptical at $z=0.2\sim 0.3$. Investigation of the image picks up at least a 
handful of more galaxies within the region marked by the X-ray contours, 
none of which
have a NIR counterpart in the 2MASS data. This is not surprising, however,
as the 2MASS nominal survey completeness limits are $m_J=15.8$, $m_H=15.1$, 
and $m_K=14.3$ mag (for pointlike sources), thus rendering the detection of 
$L<L^{\ast}$ galaxies at $z=0.256$ rather unlikely.
The two optically bright sources within $\sim 1$ arcmin East of the 
centre are stars. 

Regarding the two X-ray point sources, faint optical sources are 
present at the position of both, but neither display detectable NIR emission 
in the 2MASS data.
As will be discussed, X-ray spatial and spectral analyses suggest that the 
Western source is truly pointlike, whereas the Eastern source might contain
an extended component.
In all subsequent analysis both sources were masked out, adopting exclusion
radii of 25 arcsec and 40 arcsec for the Western and Eastern source,
respectively. These choices will be justified in \S \ref{sec,surfbright} and
\ref{sec,spectra}.

\subsubsection{XMMJ0943.9+1641}

As already mentioned, this source was not identified in {\it ROSAT} 
pointings. 
We regard the identification of the source in these data as secure, with the
innermost contour in Fig.~\ref{fig,contours} being $8\sigma$ above the 
residual background. While no clear emission peak is identifiable, 
the X-ray centroid is found to be at 
($\alpha =09^h 43^m 58\fs 5, \delta=+16\deg 41' 17''$). The structure of 
emission seems quite disturbed, but deeper 
observations are required to firmly establish its X-ray morphological 
details. 

Inspection of the 2MASS catalogue and 
Fig.~\ref{fig,contours}
revealed evidence for four NIR/optical sources within
the region covered by the displayed X-ray contours, of which only one is 
classified as extended in the infrared (roughly 0.7 arcmin from the 
X-ray centroid). None of these sources stand out as a bright central galaxy.
They all have very similar colours, consistent with those of a 
cluster population at redshifts between 0.1--0.4 (with perhaps the 
upper value being more likely, given the optical and X-ray faintness of the 
region relative to the two WARPS groups).
A second extended NIR source is present, but at a centroid distance of 
$\sim 3$ arcmin to the NW, it might rather be associated with WJ943.7.

\subsection{Surface brightness distribution}\label{sec,surfbright}

\subsubsection{One-dimensional profiles}

The radial surface brightness profiles of the two WARPS groups were initially
parametrized by fitting standard $\beta$-models, eq.~(\ref{eq,beta}), 
to the azimuthally averaged exposure-corrected emission, including a 
spatially uniform residual background. 
To minimize contamination from the 
other two extended sources in the field, exclusion regions for each of these 
were defined at their respective emission centroids
(see Fig.~\ref{fig,contours} for an example).
A residual background level was determined,
and fitting was 
done through adaptively binning the brightness profile such that each radial 
bin contained emission at $\ge$$4.5\sigma$ above this background; this 
allowed an estimate of the maximal extent of emission while at the same
time ensuring the presence of at least 20 source counts per bin (justifying
the use of $\chi^2$ statistics for fitting). 

The resulting radial profiles and best--fitting $\beta$-models are shown in 
Fig.~\ref{fig,profile}, and in Table~\ref{tab,surfbright} below we
list the derived best-fitting parameters for both groups.
Emission at $\ge$$4.5\sigma$ is detected out to at least 3 arcmin in both
cases, and the fits are acceptable, with reduced $\chi^2=0.94$ and 0.86
for 8 and 10 degrees of freedom (d.o.f.), respectively. 
Neither system shows evidence for any significant large--radius departure 
from the $\beta$-model or excess emission in the centre indicating a
cooling flow. 
Note that WJ943.5 is detected to a smaller radius than WJ943.7, although
it displays a higher central surface brightness. This reflects its larger 
$\beta$ and hence fainter fluxes at sufficiently large radii. 
For XMMJ0943.9, emission at $>4.5\sigma$ was detected out 
to $\sim 75$ arcsec but statistics were too poor to obtain useful constraints
from a $\beta$-model fit.

\begin{figure*}
\begin{center}
\mbox{\hspace{-0.0cm}
\epsfxsize=8.5cm
\epsfysize=7cm
\epsfbox{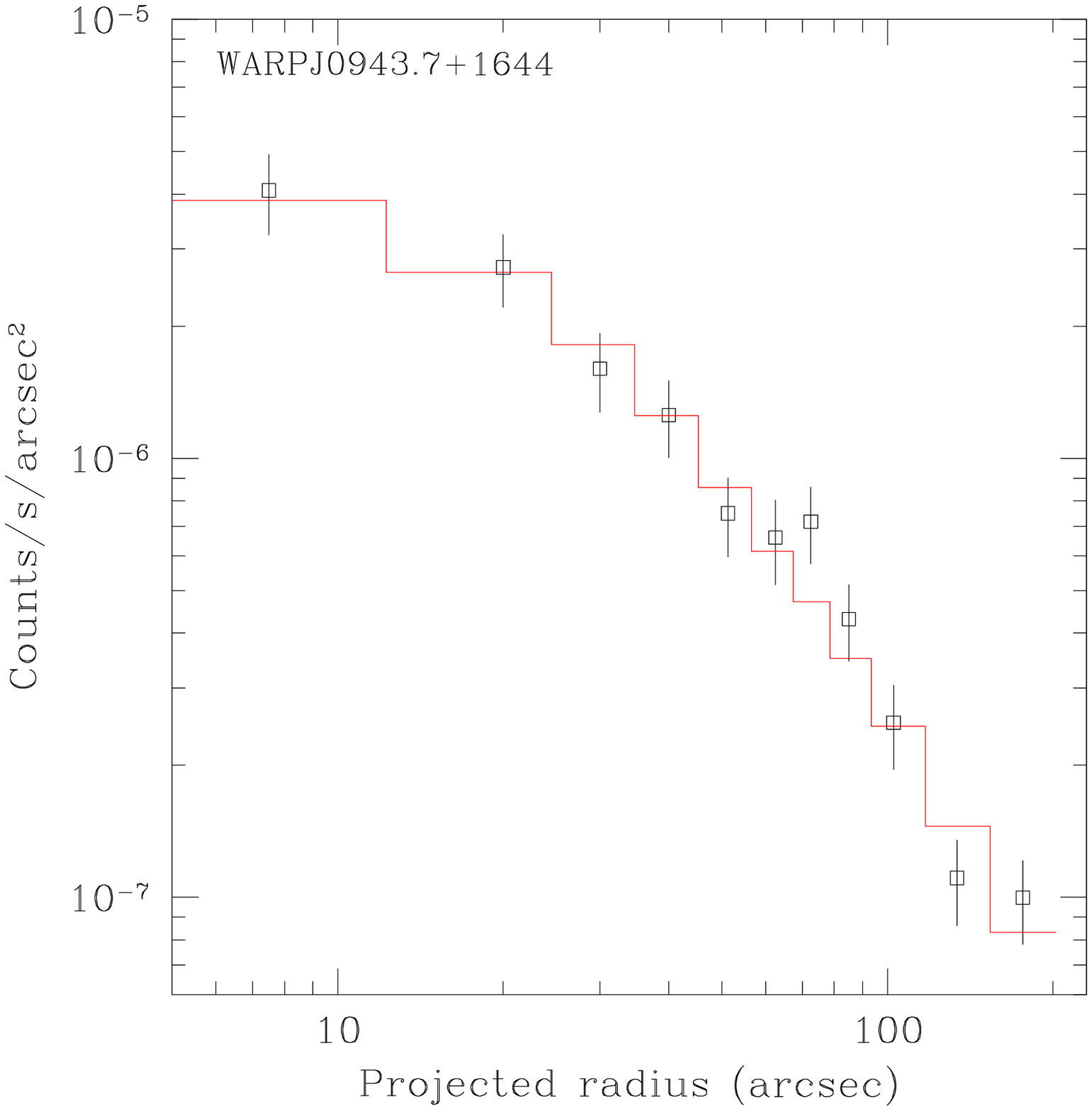}\hspace{0.5cm}
\epsfxsize=8.5cm
\epsfysize=7cm
\epsfbox{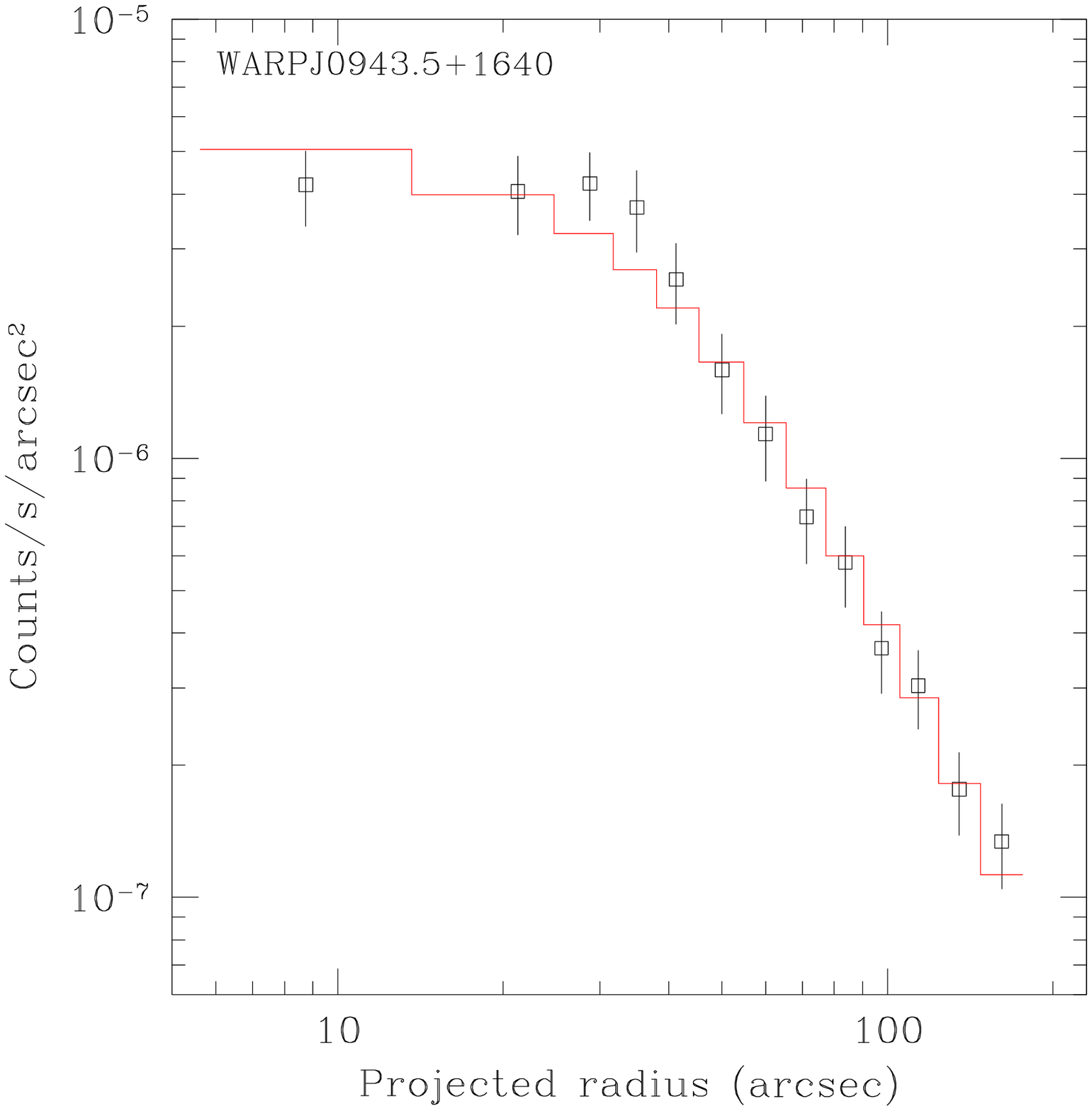}\hspace{0cm}}
\mbox{\hspace{-0.7cm}
\epsfxsize=8.5cm
\epsfysize=7cm
\epsfbox{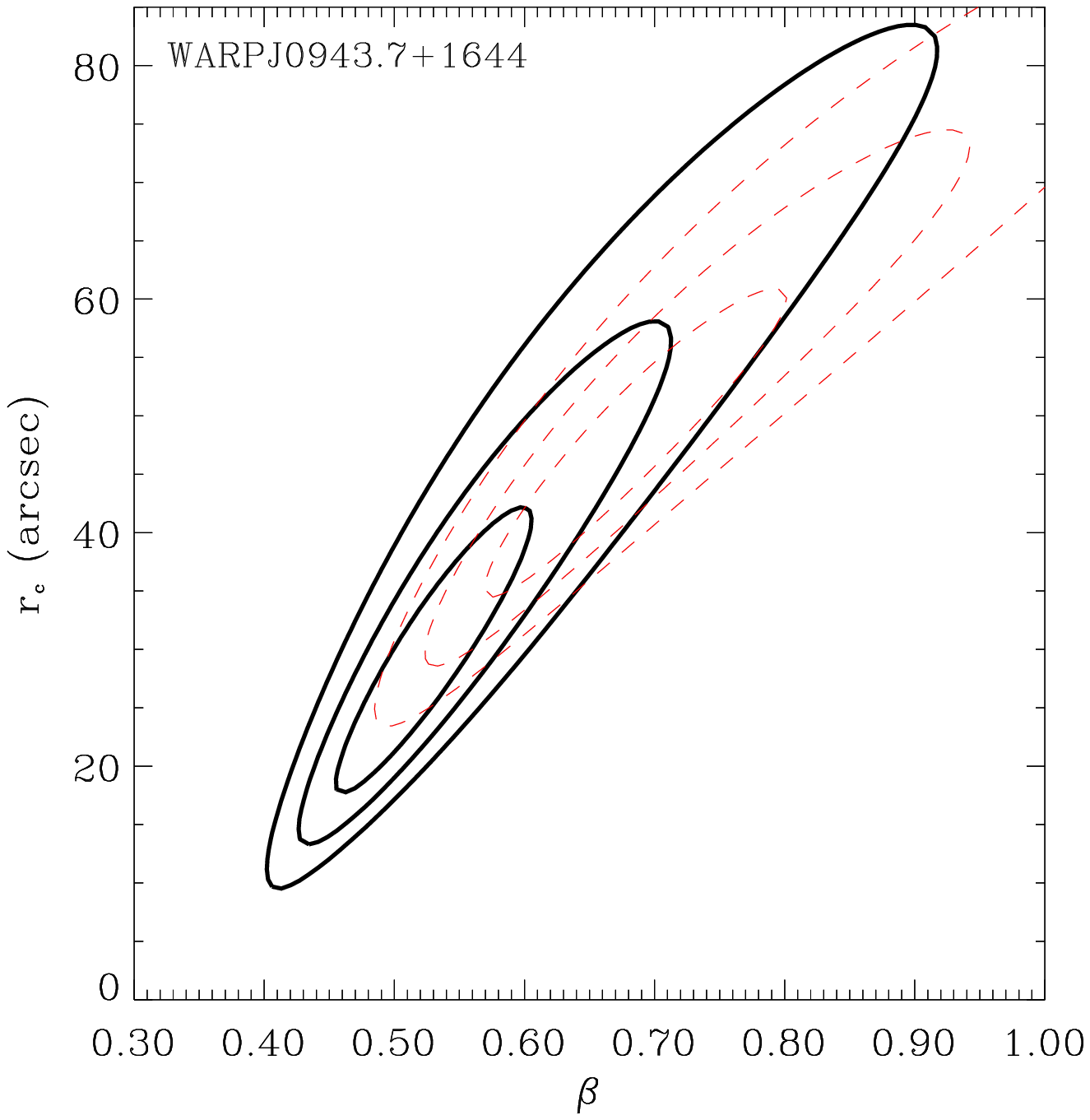}\hspace{0.5cm}
\epsfxsize=8.5cm
\epsfysize=7cm
\epsfbox{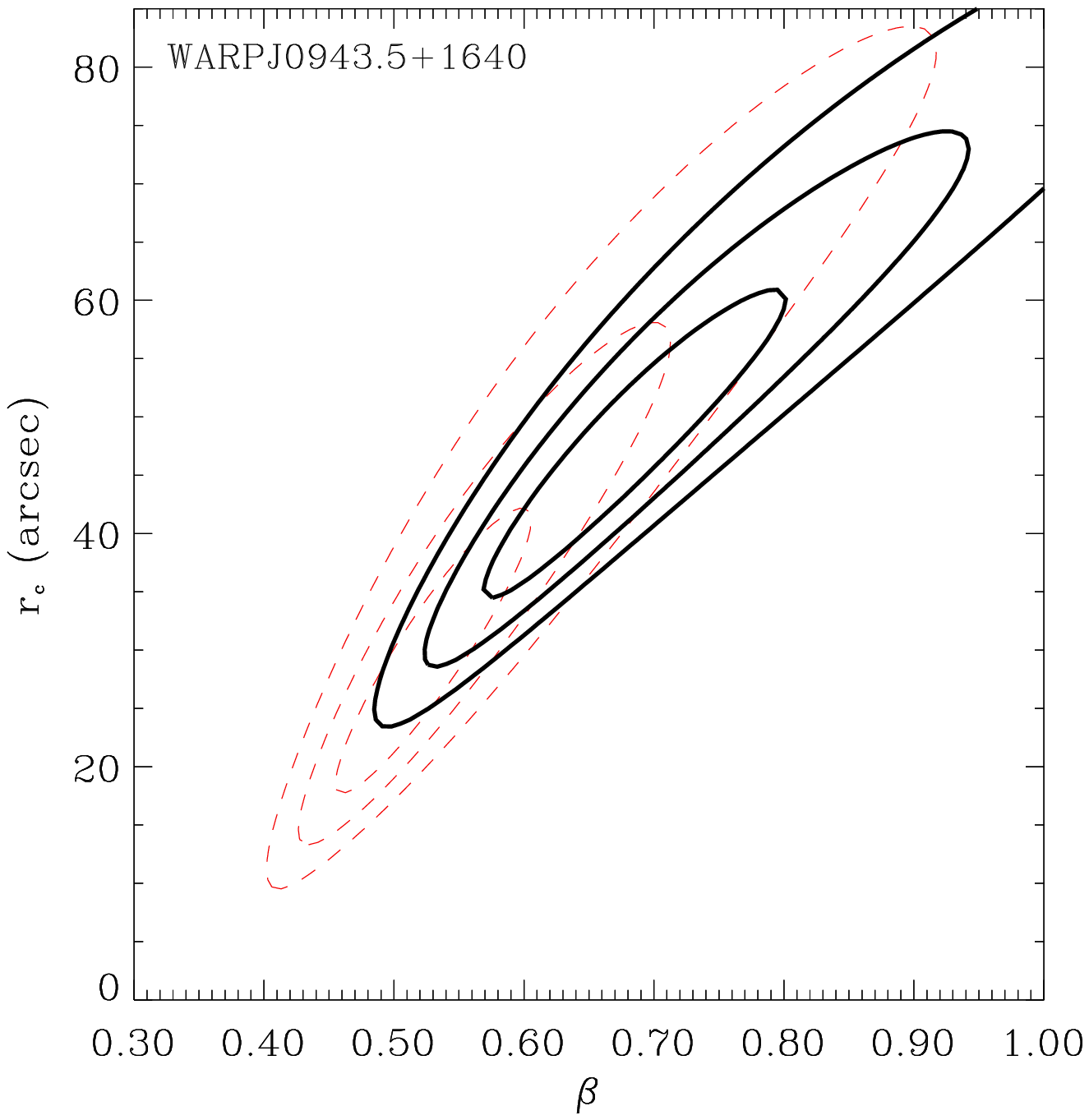}\hspace{0cm}}
\end{center}
 \caption{{\it Top}: Radial surface brightness profiles of WJ943.7 (left) 
and WJ943.5 (right), binned such that each bin contains emission at greater
than $4.5\sigma$ significance. Also shown is the best--fitting $\beta$-model
in each case. Error bars are $1\sigma$. {\it Bottom}: Corresponding 
confidence contours ($1\sigma, 2\sigma, 3\sigma$) in the $(\beta,r_c)$ plane.
Solid (dark) contours describe those pertaining to the group itself, whereas 
dashed (light) contours describe those of the other group, overlaid for
comparison. The same scales have been used in each panel for ease of
comparison.}
\label{fig,profile}
\end{figure*}

The gas density profiles in groups must break from a power law decline
at sufficiently large radii. Cosmological simulations which include
only gravity and shock heating, produce clusters in which the gas
density distribution follows that of the dark matter, and steepens
progressively towards an $r^{-3}$ dependence at large radii 
(NFW; Eke, Navarro \& Frenk 1998; Frenk et~al.\ 1999). Motivated
by this, we also fitted a gas density profile of the NFW form,
eq.~(\ref{eq,nfw}), to the surface brightness profiles. 
Since there is little evidence for
non-isothermality in the two groups (\S \ref{sec,spectra}), we assumed 
X-ray emissivity $\propto n_e^2$, where $n_e$ is the electron number
density. 
For a profile of the form of eq.~(\ref{eq,nfw}), the resulting line of sight
projection $S_{\mbox{\small NFW}}$ is not analytically tractable. Hence the 
fit was done by performing $\chi^2$--minimization on a grid of numerical 
profiles.

At large radii ($r\gg r_s$), $S_{\mbox{\small NFW}} \propto r^{-4}$, 
which is clearly much steeper than the dependence revealed by 
Fig.~\ref{fig,profile}. Large scale radii $r_s$ are 
therefore anticipated in the fits, as are large discrepancies in the inner 
regions where the observed flat density cores are incompatible with
the steep NFW behaviour. 
Consequently, the data points inside the best-fitting core radii of the 
$\beta$-models were excluded in the fits. 
For both groups, the isothermal NFW fits are good, yielding fit qualities 
comparable to
those obtained using $\beta$-models, the fit being slightly better for 
WJ943.5 and slightly worse for WJ943.7 than a $\beta$-model fit.
Thus, assuming isothermal gas, the data are in accordance with models where 
the gas density shows a progressively steepening power-law index, as it does 
for an NFW model.
As will be shown in \S \ref{sec,noniso}, the X-ray emissivity of the groups
is independent of temperature to within $\sim5$ per cent for the relevant 
range of parameters involved, so the above result applies even if the gas is 
moderately non-isothermal.
However, the scale radii obtained are in all cases large: for WJ943.5 
$r_s \sim 630 h^{-1}$ kpc and for WJ943.7 $r_s \sim 850 h^{-1}$ kpc 
(even larger than the derived $r_{200}$, cf.\ \S ~\ref{sec,mass}). This
corresponds to small concentration parameters $c = r_{200}/r_s < 1$, and is
in apparent conflict with
expectations from numerical simulations (NFW; Eke et~al.\ 1998; 
Moore et~al.\ 1998; Lewis et~al.\ 2000) and 
other observations (e.g.\ Pratt \& Arnaud 2003) which typically find 
$r_s\simeq 0.1-0.25r_{200}$ (i.e.\ $c \approx 4-10$) for the dark matter 
distribution.
Given that we are dealing here with the {\it gas} density, this most
likely simply reflects the fact that the gas distribution
is much less centrally concentrated than that of the dark matter, as will be 
substantiated in \S \ref{sec,discus}. Models in which gas traces total mass 
are hence clearly ruled out for both our groups.
This is also supported by the results of
Ettori \& Fabian (1999), who use a similar approach for a sample of massive 
clusters, and find a mean ratio between NFW gas scale radius $r_s$ and 
$\beta$-model core radius $r_c$ of $r_s/r_c \approx 3$, whereas we find 
$r_s/r_c \ga 15$ for WJ943.7 and $r_s/r_c \ga 5 $ for WJ943.5.
Importantly, the large scale radii resulting from the fits also show that the
derived surface brightness profiles are inconsistent with a steepening in the
logarithmic density slope
$\eta = - (\mbox{d}\, \mbox{{ln}}\, n_e / \mbox{d}\, \mbox{{ln}}\, r)$
towards $\eta \approx3$ at large radii. In fact, the best-fitting 
NFW models produce slopes of $\eta=1.66$ and 1.80 for WJ943.7 
and $\eta=1.87$ and 2.08 for WJ943.5 at the derived extent of emission 
$r_{ext}$ (see Table~\ref{tab,surfbright} below) and at $r_{200}$ 
(\S \ref{sec,mass}), respectively. 
For both groups, the concluding point is therefore that isothermal NFW models
only provide a good description of the data if 
the associated scale radii are so large as to make the resulting surface 
brightness profiles rather similar to $\beta$-models.

\begin{table*}
\begin{minipage}{178mm}
\caption{Best-fitting values from 1-D and 2-D surface brightness fitting 
(0.4--2.5 keV). 
All uncertainties are 1$\sigma$ for one interesting parameter. 
Physical scales are computed for
distances corresponding to the spectroscopic redshifts of the groups. Radial
extents $r_{ext}$ specify the outer radius of the outermost radial bin 
containing a $4.5\sigma$ detection. Position angles are given 
counter-clockwise from the North. The goodness of fit for 2-D models is as 
described in the text.}
\label{tab,surfbright}
\begin{center}

{\bf 1-D}

\begin{tabular}{lcccccccc}
\hline
Source & Redshift & $S_0$ & $\beta$ & $r_c$ & $r_c$ & $\chi^2 /$d.o.f.\ & 
$r_{ext}$ & $r_{ext}$ \\
 & & (cts s$^{-1}$ arcsec$^{-2}$)  &  & (arcsec) & ($h^{-1}$ kpc) & & (arcsec) & ($h^{-1}$ kpc) \\ \hline
\vspace{.15cm}
WJ943.7 & 0.180 & $4.18^{+1.02}_{-0.86} \times 10^{-6}$ & 0.51$^{+0.05}_{-0.04}$ & 27.0$^{+9.1}_{-6.5}$ & 57$^{+19}_{-17}$ & 7.5/8 & 200 & 425 \\
WJ943.5 & 0.256 & $5.33^{+0.72}_{-0.67} \times 10^{-6}$ & 0.66$^{+0.08}_{-0.06}$ & 45.7$^{+9.4}_{-7.7}$ & 127$^{+26}_{-21}$ & 8.6/10 & 175 & 490 \\  \hline
\end{tabular}

\vspace{0.3cm}
{\bf 2-D}

\begin{tabular}{lccccccc}
 \hline
Source & $S_0$ & $\beta$ & $r_c$ & $r_c$ & $e$ & $\theta$ & Goodness \\
       &   (cts s$^{-1}$ arcsec$^{-2}$) &  & (arcsec) & ($h^{-1}$ kpc) & & & of fit\\ \hline
\vspace{.15cm}
WJ943.7 & $4.60^{+0.56}_{-0.50} \times 10^{-6}$ & 0.49$^{+0.03}_{-0.03}$ & 24.9$^{+4.2}_{-3.6}$ & 53$^{+9}_{-8}$ & 0.00$^{+0.07}_{-0.00}$ & -- & 0.93 \\
WJ943.5 & $5.93^{+0.37}_{-0.68} \times 10^{-6}$ & 0.63$^{+0.08}_{-0.03}$ & 45.2$^{+12.6}_{-4.1}$ & 126$^{+35}_{-11}$ & 0.19$^{+0.09}_{-0.04}$ & $81^{\circ}$$^{+8}_{-9}$ & 0.16 \\ \hline
\end{tabular}
\end{center}

\vspace{.3cm}

\end{minipage}
\end{table*}

\subsubsection{Two-dimensional profiles}

Although the X-ray emission of both WARPS groups is reasonably symmetric and 
is well fitted by a one-dimensional $\beta$-profile, two-dimensional
$\beta$-models were also fitted to the emission, using {\sc sherpa}, 
in order to further investigate
the surface brightness structure. 
This approach has the 
advantage that the effects of the asymmetric EPIC PSF can be accurately 
taken into account in the model fit.
However, whilst 2-D fitting takes better advantage of the 
available spatial information, the image counts do not satisfy the
requirement of Gaussian errors which underlies the $\chi^2$ 
statistic (the number of counts
in some data bins may be very low). Following Helsdon \& Ponman (2000), 
the maximum--likelihood based Cash statistic (Cash 1979) was used instead.
Vignetting was taken directly into account in
the fit through including the exposure map in the modelling of the source.

The 2-D fit results are also presented in Table~\ref{tab,surfbright}. 
For each source, the PSF model employed for convolution with the source model
was produced at the relevant detector position at
a photon energy of 1 keV using {\sc xmmsas}.
Additional free parameters 
compared to the 1-D fit were the eccentricity 
$e \equiv (1-b^2/a^2)^{1/2}$, position angle $\theta$, and 
$x$- and $y$-centre of the X-ray gas distribution. 

No estimator of the goodness of fit is
immediately available when using the Cash statistic.
Hence to address the issue of fit quality, we
adopted a Monte Carlo approach, generating for each source 1000
artificial images based on the best-fitting model. In each case, Poisson 
noise was added to the input model image, which was subsequently convolved 
with the PSF model and vignetted. Model fits were then performed to these 
images in the same manner as for the real data, recording for each fit the 
value of the Cash statistic, and finally fitting a Gaussian to the resulting 
distribution of these values. Comparing the Cash statistic, $C$, derived for 
the real data to this distribution, a measure of the goodness of fit was
obtained as the number of standard deviations $\sigma$ separating $C$ from 
the centre of the distribution. This value is also presented in 
Table~\ref{tab,surfbright}, and it is seen that both fits are reasonably 
good, being less than $1\sigma$ from the distribution centre. 
Note also that confidence intervals can be assessed as for $\chi^2$ 
statistics, since
{\em differences} in the Cash statistic are $\chi^2$--distributed.

For both groups, the 1-D and 2-D results agree  well within the errors.
%
For WJ943.7, the derived $\beta$--value is typical for groups 
and poor clusters with X-ray temperatures below 1--2 keV.
The X-ray morphology is consistent with zero eccentricity, 
and $\theta$ is therefore unconstrained for this source.
WJ943.5, in contrast, appears to display a larger X-ray core and a 
$\beta$--value more typical of richer clusters ($T\ga 3$ keV). 
The derived eccentricity of its X-ray emission, 
$e \approx 0.2$, is consistent with typical values found for other 
groups 
and clusters (Mohr et al.\ 1995; Helsdon \& Ponman 2000). {\it ROSAT} 
results for other groups (Mulchaey \& Zabludoff 1998) have shown that the
optical light of the brightest group/cluster galaxy is often well aligned
with the position angle of the X-ray emission. In the case of WJ943.5, 
however, the X-ray position angle $\theta_X \sim 80\deg$ seems not 
particularly well aligned with the $R$-band optical light of the central 
galaxy, which shows $\theta_{opt} \sim 50\deg$. In the central 
regions, within 1 arcmin of the emission peak, the
X-ray contours are nearly circular ($e \approx 0.02$), but the 
best-fitting position angle of $\theta_X \sim 20\deg$ is effectively 
unconstrained and so is consistent with both of the above values. 

The difference in $r_c$, and particularly $\beta$, seen for the two groups is
interesting, inasmuch as the systems have reasonably similar
temperatures (\S \ref{sec,spectra}) and hence masses. 
To investigate the significance of the difference, we also plot in 
Fig.~\ref{fig,profile} the derived confidence contours in the 
$(\beta,r_c)$ plane for both groups. Although the $1\sigma$ contours 
marginally overlap, it is clear that the two profiles are not consistent
at $1\sigma$. For WJ943.5, any combination of $r_c$ and $\beta$ leading to
a typical group value of $\beta \la 0.5$ is only marginally allowed within
$3\sigma$ confidence. It could of course be conjectured that some
process has lead to a significantly larger X-ray core for this group, which,
given the quality of the data, naturally forces $\beta$ towards high values 
in the fits. 
Another possibility is related to the fact that
when masking out the Eastern point source seen in 
Fig.~\ref{fig,contours} to the standard radius of 25 arcsec, resulting
residuals from 2-D surface brightness fitting suggest the source
is not quite pointlike 
(whereas the W source is well accounted for by point source emission). This
was confirmed by explicitly including, in the overall fit to the group 
emission, a point source model at the position of the E source, the result 
still showing systematic residuals around the source. 
The results
listed in Table~\ref{tab,surfbright} are therefore based on an exclusion 
radius of 40 arcsec for this source, estimated from the 2-D fitting residuals
to mask out the excess emission to sufficient accuracy.
Furthermore, as will be described below, spectral
results also suggest that the Eastern source is not pointlike. If indeed a
'separate' extended component is present at $\sim$ 1 arcmin East of the
group emission centroid, this might also drive $r_c$, and hence $\beta$,
towards high values in the fits. 

We note that an eccentricity of $\sim0.2$ persists when explicitly
including the Eastern source in the 2-D modelling, either as pointlike or
extended. This strongly suggests that the derived elongation and position 
angle of the group X-ray contours are intrinsic and not
caused by excess emission from an extended source embedded in the group
emission.

\subsection{X-ray temperatures}\label{sec,spectra}

For the spectral analysis, integrated spectra of the WARPS groups out to 
$r_{ext}$ were extracted for each instrument and observation,
accumulating the spectra in bins containing at least 20 counts.
For each source, all six background--subtracted spectra
were then fitted simultaneously in the 0.4--4 keV band,
using an absorbed {\em mekal} model in {\sc xspec} v11.1, with the
redshift fixed at the spectroscopic value and $N_H$ fixed at 
the Galactic value of $3.2\times 10^{20}$ cm$^{-2}$ (Stark et al.\ 1992).
Free parameters were thus temperature, metal abundance, and spectral 
normalization. Results are presented in Table~\ref{tab,spectra}, with
metallicities derived using the solar abundance table of Anders \& Grevesse
(1989).
Acceptable fits were obtained, with best-fitting temperatures and abundances 
characteristic of X-ray bright groups or poor clusters.

For WJ943.5 it was found that the spectrum could only be adequately fitted
by one thermal component, 
provided the Eastern point source was excised out to at least 40 arcsec, 
once again suggesting that this source is extended rather than pointlike.
The combined spectrum of the two 'point' sources  seen in the group 
is nevertheless well
modelled by a power-law of slope $1.8\pm 0.2$ added on top of the 
thermal plasma spectrum of the intragroup medium.
However, masking out the Western source and modelling the Eastern alone 
along with a thermal plasma representing the group emission, we find that 
in all cases marginally better fits are obtained when replacing the 
power--law model with another {\em mekal} thermal plasma at the group 
redshift, fixed at a metallicity of $0.3$Z$_{\odot}$. 
We find no evidence
that the source is rather described by multiple power-laws or a single
power-law with internal absorption. Allowing the
redshift of the {\it mekal} model to vary yields $z$ in the interval 
0.2--0.3, i.e.\ consistent with the group redshift.
These conclusions apply independently of the exact choice of fitting region,
and seem to suggest the presence of a separate blob of 
plasma at a temperature roughly 0.5 keV hotter than the surrounding medium. 
Due to the limits on the data, a hardness map does not provide any additional
information in this context.

For XMMJ0943.9, we also attempted to fit a thermal plasma model to 
the 0--75 arcsec spectrum, keeping temperature, 
redshift, and normalization as free parameters, while fixing the abundance at 
$0.3$Z$_{\odot}$. A {\it mekal} model provides a good fit 
($\chi^2=20.9$ for 23 d.o.f.), yielding
$T=1.40^{+0.22}_{-0.18}$ ($1\sigma$ errors) and a redshift of 
$z=0.33^{+0.06}_{-0.07}$ (90 per cent confidence). With both parameters 
reasonably well determined, this suggests that XMMJ0943.9
is a yet more distant galaxy group. The derived redshift is in good
agreement with that expected from the optical-NIR colours of sources in
this region, cf.\ \S \ref{sec,morph}. We derive an unabsorbed 0.5--2 keV
flux of $\sim 3\times 10^{-14}$ ergs cm$^{-2}$ s$^{-1}$, corresponding to a
luminosity of $1.0\times 10^{43} h^{-2}$ ergs s$^{-1}$ at the nominal 
best-fitting redshift.

\begin{table}
\caption{Results of {\em mekal} fits to the integrated 0.4--4 keV spectra, 
with $N_H$ fixed at the 
Galactic value of $3.2\times 10^{20}$ cm$^{-2}$. Luminosities are given in
the 0.5--2 keV band in units of $10^{43} h^{-2}$ ergs s$^{-1}$ (see text
for details).
Errors are $1\sigma$ for one interesting parameter.}
\label{tab,spectra}
\begin{tabular}{lcccc}
\hline
Source & $T$/keV & $Z/$Z$_{\odot}$ & $\chi^2$/dof & $L_{X}$  \\ 
 & & & & ($10^{43}$ erg/s) \\ \hline
\vspace{.15cm}
WJ943.7 & $1.67^{+0.12}_{-0.15}$ & $0.34^{+0.12}_{-0.11}$ & 164/168 & 
$0.57^{+0.07}_{-0.07} $\\ 
\vspace{.5cm}
WJ943.5 & $2.40^{+0.40}_{-0.39}$ & $0.31^{+0.24}_{-0.16}$ & 126/122 &
$1.97^{+0.25}_{-0.25}$
\vspace{-.5cm} \\ \hline
\end{tabular}
\end{table}

Due to the limited statistics available (we record $\sim$1500 and $\sim$1100 
background-subtracted source counts for WJ943.7 and WJ943.5, 
respectively, within the non-excluded 
regions), a detailed study of temperature and abundance gradients in
the intra-cluster medium (ICM) is not feasible. Although a hardness map does 
not reveal evidence for any significant temperature variations within either 
group, an examination of the 
gross structure of the ICM temperature distribution $T(r)$ in the two groups
was nevertheless attempted by extracting spectra in two or three concentric 
annuli, containing roughly equal numbers of source counts. 
The spectra were then fitted as above, with abundances fixed at the value 
derived from the fits to the integrated spectra inside $r_{ext}$ 
(consistent results for both $T$ and $Z$ were obtained when the 
latter was allowed to vary).
In all bins good fits are obtained, and results for two bins are shown in 
Fig.~\ref{fig,tprof}. Although both groups show tentative evidence for a 
mild ($\sim$$0.5-1$ keV) temperature decline with radius, 
the ICM is consistent with being isothermal 
within the 90 per cent confidence ($1.65\sigma$) errors in each bin. 
Specifically, there is 
no indication of the presence of cool cores in the central regions. This
is consistent with the lack of excess emission in the group cores 
relative to the best-fitting isothermal $\beta$-model.
When spectral properties are derived using three rather than 
two annuli, these are, of course, subject to larger
statistical uncertainties, but are again consistent with isothermality, 
and show no central cooling.

\begin{figure*}
\begin{center}
\mbox{\hspace{-0.9cm}
\epsfxsize=8.9cm
\epsfysize=7cm
\epsfbox{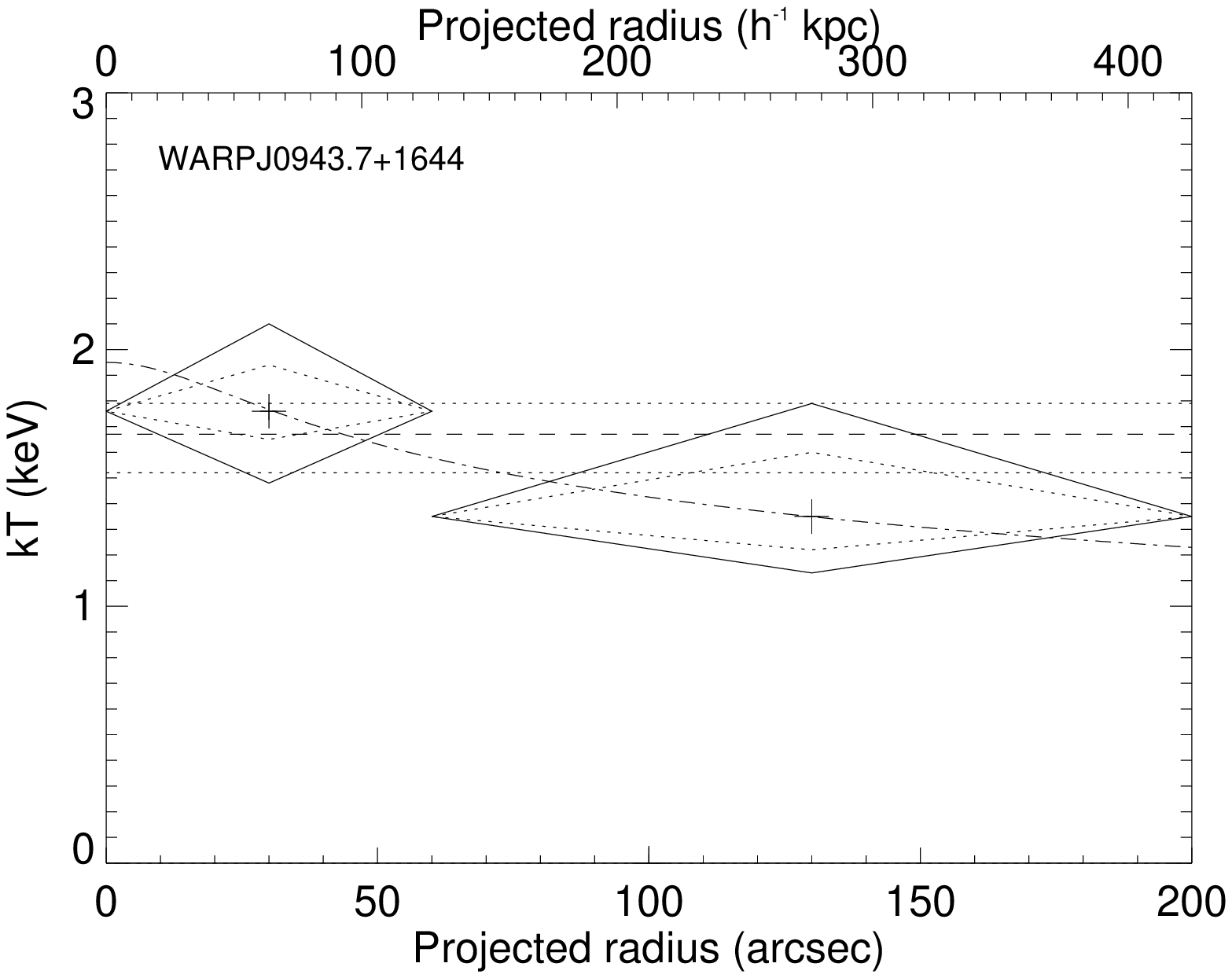}\hspace{0.4cm}
\epsfxsize=9.3cm
\epsfysize=7cm
\epsfbox{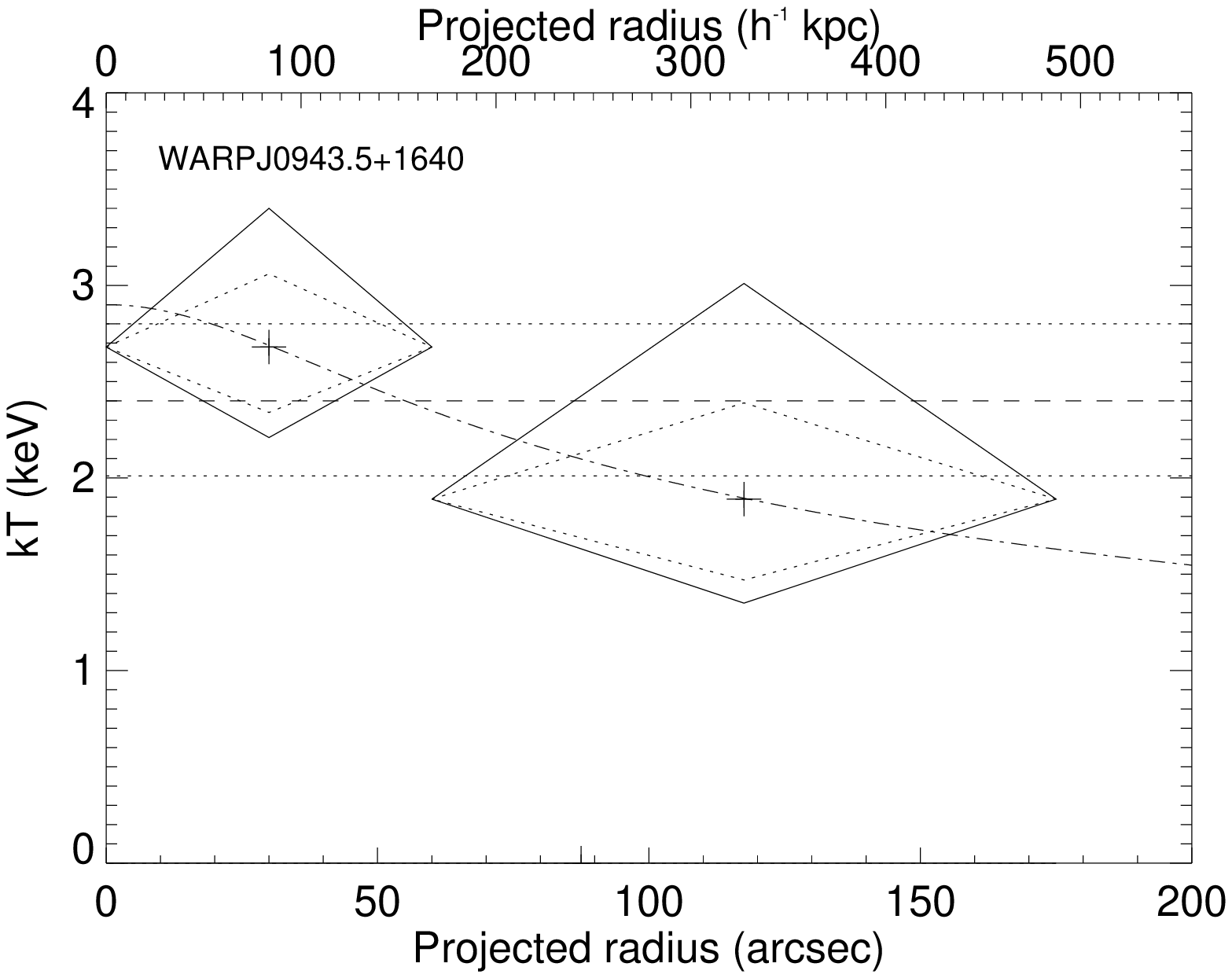}\hspace{0cm}}
\end{center}
 \caption{Projected radial variation of the X-ray gas temperature in
WJ943.7 (left) and WJ943.5 (right). Crosses mark the best-fitting values
in each bin, with surrounding solid (dotted) lines showing the 
90 per cent ($1\sigma$)  confidence
intervals. Dashed horizontal lines show the best-fitting value from the 
integrated spectrum, with the associated $1\sigma$ errors marked by dotted 
lines. The dot-dashed curves represent the polytropic models discussed in
the text.}
\label{fig,tprof}
\end{figure*}

\subsection{X-ray fluxes and luminosities}\label{sec,fluxes}

From the best-fitting {\em mekal} models for the integrated spectra,
resulting 0.5--2 keV unabsorbed fluxes for 
WJ943.7 (WJ943.5) are 
$0.96^{+0.11}_{-0.12} 
(1.12^{+0.14}_{-0.14})\times 10^{-13}$ ergs cm$^{-2}$ s$^{-1}$
within the regions adopted for spectral fitting. Errors were determined from
the fractional errors on the spectral normalisation.
Correcting the WJ943.5 flux inside the spectral region for removed 
area, assuming the nominal best-fitting 2-D $\beta$-model in 
Table~\ref{tab,surfbright}, the resulting flux is 
$1.43^{+0.19}_{-0.18}\times 10^{-13}$ ergs cm$^{-2}$ s$^{-1}$, 
where the quoted $1\sigma$ errors are from the spectral flux determination 
alone. 



At the relevant redshifts, the derived fluxes within $r_{ext}$ 
correspond to unabsorbed 0.5--2 keV restframe luminosities of 
$L_X=0.57^{+0.07}_{-0.07}\times 10^{43} h^{-2}$ ergs s$^{-1}$
(WJ943.7) and $1.97^{+0.25}_{-0.25}\times 10^{43}$ $h^{-2}$ ergs s$^{-1}$
(WJ943.5, corrected for removed area as above), as listed in 
Table~\ref{tab,spectra}.
In Fig.~\ref{fig,lxt} the best-fitting $T_X$ and the bolometric 
(0.01--15 keV rest-frame) luminosities within the radii adopted for spectral
fitting are compared to
those of the low--redshift ($z < 0.04$) $T\geq 1$ keV groups in the samples 
of Helsdon \& Ponman (2000) and Xue \& Wu (2000). Note that both 
depicted $L_X-T$ relations for groups 
($L_X\propto T^{4.9\pm 0.8}$, Helsdon \& Ponman 2000; $L_X\propto T^{5.57\pm1.79}$, Xue \& Wu 2000) have been derived using groups at 'all' $T$,
whereas only groups with $T \ge 1$ keV are included in Fig.~\ref{fig,lxt}.
Note also that while the groups in the
Helsdon \& Ponman sample were analysed in a coherent manner,
data for the Xue and Wu groups were compiled from a number of 
different sources. In neither case was the X-ray emission of the groups 
derived within a consistent radius, be it physical or as a fraction of 
$r_{200}$, and thus a rigorous comparison to our results cannot be made.
Nevertheless, it is found that both WARPS groups have luminosities consistent
with those derived for other poor clusters, agreeing well with the cluster 
$L_X-T$ relation of slope $2.79\pm0.08$ derived by Xue \& Wu (2000). 
This $L_X-T$ relation is based on a sample of 
274 clusters drawn from the literature, the large majority of which are at
$z<0.5$, and agrees well at the relevant temperatures 
($1\la T \la 3$ keV), with, e.g., the 
low-to-moderate redshift ($z<0.2$) relation derived for the $T>1$ keV WARPS 
clusters (Fairley et~al.\ 2000) and with estimates for $z<0.1$ clusters
from {\it BeppoSAX} data (Ettori, De Grandi \& Molendi 2002).

\begin{figure}
\hspace{-0.7cm}
\epsfxsize=9.2cm
\epsfysize=7.5cm
\epsfbox{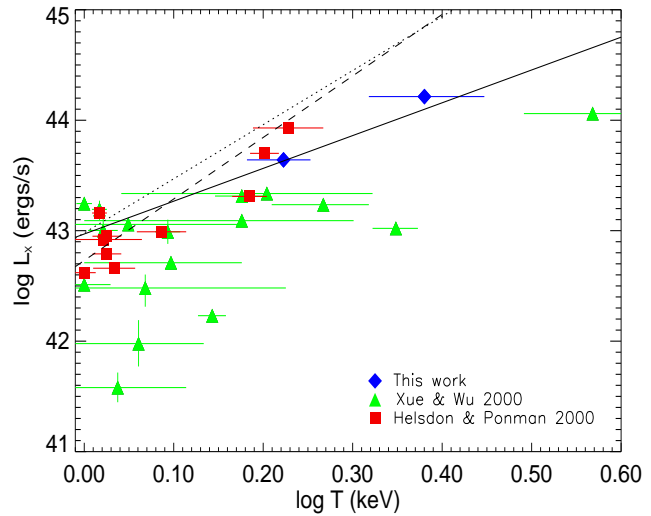}
\caption{The derived bolometric luminosities and temperatures of the two
WARPS groups (diamonds), shown together with the values for the 
$T\geq 1$ keV
groups of Xue \& Wu (2000; triangles) and Helsdon \& Ponman (2000; squares). 
Also shown is the nominal best-fitting $L_X-T$ relation for {\em all}
groups in the samples of Xue \& Wu (dashed line), Helsdon \& Ponman (dotted
line), and for the cluster sample of Xue \& Wu (solid line).
The figure assumes an Einstein-de Sitter cosmology with $h=0.5$.}
\label{fig,lxt}
\end{figure}

\section{Mass analysis}\label{sec,mass}

\subsection{Effects of non-isothermality}\label{sec,noniso}

It is clear from Fig.~\ref{fig,tprof} that the temperature profiles are not 
resolved with sufficient accuracy to immediately allow detailed profiles to 
be made of 
temperature-based quantities such as mass or gas mass fraction. 
Fig.~\ref{fig,tprof} and the ensuing discussion shows, however, that there is
no evidence for significant departures from isothermality in either group
at any radius probed,
and an isothermal distribution is, indeed, the flattest
reasonable temperature distribution with which the data are consistent. 
One might argue that $T$ in the innermost bin could have 
been underestimated due to the presence of a cooling flow;
this would produce a steeper temperature profile outside the cooling core 
and hence a larger deviation from
isothermality than is immediately apparent in Fig.~\ref{fig,tprof}. While 
the expected central cooling times (using the cooling curves of Sutherland 
\& Dopita 1993, for a $Z=0.3Z_{\odot}$ plasma) of $\sim 7$ and $\sim 11$ Gyr 
for WJ943.7 and WJ943.5, respectively, may suggest the presence of weak 
cooling flows, this is not supported by the derived surface brightness 
profiles.
Moreover, a number of recent studies of low-temperature systems with 
well-defined
temperature profiles have demonstrated that, outside a possible cool core, 
isothermality may indeed be a good assumption (e.g.\ Mushotzky et~al.\ 2003;
Pratt \& Arnaud 2003).
In the following sections we will therefore assume the ICM to be isothermal
at the mean temperature derived inside $r_{ext}$.
It is nevertheless instructive to consider the impact of 
assuming isothermality, in order to assess the potential error in doing so,
and in all subsequent analysis we will attempt to quantify potential effects 
of this assumption on our results. 

On the basis of the temperature data of Fig.~\ref{fig,tprof}, one can 
assume model profiles for $T(r)$ which 
(i) are consistent with the data, (ii) can be justified
with reference to other observed and simulated systems, and (iii) can 
provide an indication of the systematic errors associated with making a
single isothermal assumption for $T(r)$. Hence, in the absence of detailed 
temperature data, such model profiles are used in the following in conjunction
with the isothermal assumption. We readily acknowledge the fact that the 
resulting radial variation of various quantities derived from $T(r)$ is, 
strictly speaking, modelled rather than observed, providing an approximation 
to the true profile which is only reliable to the extent to which the ICM 
temperature behaves as assumed. As will be seen, the overall conclusions
are nevertheless not sensitive to specific choices for $T(r)$ allowed by
Fig.~\ref{fig,tprof}.

The detailed shape of the temperature profile in our groups is largely 
unconstrained. There is, however, compelling evidence from both observations 
and simulations of groups and clusters that a polytropic equation of state,
\begin{equation}
\label{eq,poly}
T\propto \rho ^{\gamma -1} ,
\end{equation}
provides a good description of any radial decline in gas temperature outside a
possible cool core. This is the case both for simulated clusters
(Lewis et~al.\ 2000; Loken et~al.\ 2002), simulated groups/cool clusters 
(Ascasibar et~al.\ 2003),
and for observed clusters with well-determined temperature profiles 
(e.g.\ Markevitch et~al.\ 1998; Ettori \& Fabian 1999; 
Markevitch et~al.\ 1999; Pratt, Arnaud \& Aghanim 2001; Pratt \& Arnaud 2002)
or surface brightness profiles (Miralda-Escude \& 
Babul 1995). It is clearly also a better description than the isothermal 
assumption in some clusters with less well-determined temperature 
profiles (see e.g.\ Sarazin, Wise \& Markevitch 1998).
Eq.~(\ref{eq,poly}) is consequently employed here as a tool to investigate
the systematic errors associated with assuming isothermality.
For the 
temperature data shown in Fig.~\ref{fig,tprof}, 
polytropic models intercepting the nominal best-fitting temperature in each 
bin have indices of 
$\gamma=1.15$ for WJ943.7 and $\gamma=1.22$ for WJ943.5, very similar to 
values found for observed (Markevitch et~al.\ 1999; Ettori \& Fabian 1999) 
and simulated (Lewis et~al.\ 2000; Ascasibar et~al.\ 2003) clusters. 
Here we shall not attempt to include errors on 
these indices; given the available photon statistics, they would not be well 
constrained for the present data and their errors would propagate into large
uncertainties on the resulting total masses.

If indeed the gas is polytropic, i.e.\ satisfying eq.~(\ref{eq,poly}), and if
$\rho$ follows a $\beta$-model, the true gas temperature $T_{true}$ will 
differ from the projected (i.e.\ measured) one $T_{proj}$ only by a constant 
factor,
\begin{equation}
\label{eq,ttrue}
\frac{T_{true}}{T_{proj}} =  \frac{ \Gamma[\frac{3}{2}\beta(1+\gamma)]\Gamma(3\beta-\frac{1}{2})}
{\Gamma[\frac{3}{2}\beta(1+\gamma)-\frac{1}{2}]\Gamma(3\beta) }
\end{equation}
(Markevitch et al.~1999), which amounts to 1.07--1.08 for the systems studied
here. Strictly speaking, eq.~(\ref{eq,ttrue}) is only valid assuming 
emissivity $\propto n_e^2 T^{\alpha}$,
but in the 0.4--4 keV band considered for spectral fits, and over the
relevant temperature ranges (1.3--2.0 keV for WJ943.7 and 1.6--2.9 keV for 
WJ943.5; see Fig.~\ref{fig,tprof}), 
the emissivity is temperature--independent to within 5 per cent and is 
$\propto T^{\alpha}$ to within 3 per cent for a 
$Z=0.3$Z${\odot}$ {\it mekal} plasma.

\subsection{Total masses and virial radii}

Motivated by the regularity of the X-ray emission in the groups, we assume 
spherical symmetry and hydrostatic equilibrium of the X-ray emitting gas.
The total gravitating mass inside a radius $r$ is then given by
\begin{equation}
M_{grav}(<r)=-\frac{kT(r)r}{G\mu m_p} \left(\frac{\mbox{d}\, \mbox{ln}\,
\rho_{gas}}{\mbox{d}\, \mbox{ln}\, r} +
\frac{\mbox{d}\, \mbox{ln}\, T}{\mbox{d}\, \mbox{ln}\, r}
 \right),
\label{eq,mass}
\end{equation}
where $\mu$ is the mean molecular weight and $m_p$ is the proton mass.
Numerical simulations (Evrard, Metzler \& Navarro 1996) show that this is a 
reliable mass estimator out to at least $r_{500}$, the radius containing a
mean overdensity of 500 with respect to the critical 
density $\rho_c(z)$, which is comparable to the extent of emission 
$r_{ext}$ in both groups (see below). Taking $\mu=0.6$ and
using the emission-weighted mean temperatures given in 
Table~\ref{tab,spectra}, along with the derived parameters for the gas 
profiles, the resulting total masses within $r_{ext}$ are 
presented in Table~\ref{tab,mass}. Adopting a Monte Carlo approach, the 
associated uncertainties were estimated by constructing 10,000 artificial 
mass profiles. For each of these, the values of involved parameters 
($r_c$, $\beta$, $T$) were taken from an asymmetric Gaussian distribution 
centered at the best-fitting value and having characteristic upper and lower 
widths given by the 
associated $1\sigma$ uncertainties (Tables~\ref{tab,surfbright} and 
\ref{tab,spectra}). The derived distribution of masses was in general also an
asymmetric Gaussian;
the quoted mass errors are the associated standard deviations 
(upper and lower) with respect to the nominal best-fitting values.

Assuming the groups to be virialized out to $r_{200}$, the virial radius can 
be self-consistently computed from the derived mass profile once
$\rho_c(z)$ is known. Using
\begin{equation}
\rho_c(z) = \frac{3H_0^2}{8\pi G}[(1+z)^3 \Omega_m + (1+z)^2(1-\Omega_m-\Omega_{\Lambda})+\Omega_{\Lambda}],
\end{equation}
where the term in brackets is the ratio $(H(z)/H_0)^2$, we derive nominal 
virial radii $r_{vir} = r_{200}= 0.57 h^{-1}$ and 
0.74$h^{-1}$ Mpc for WJ943.7 and WJ943.5, 
respectively, in the adopted $\Lambda$CDM cosmology. We list $r_{200}$ and
$r_{500}$ relative to $r_{ext}$ in Table~\ref{tab,mass}; 
it is seen that emission has been detected beyond $r_{500}$ and out to 
74 and 66 per cent of $r_{200}$ in the two cases.


A temperature gradient in the
X-ray gas changes total masses, and hence also $r_{200}$ and $r_{500}$.
Assuming the above polytropic gas profiles lowers the estimated masses
within $r_{500}$ and $r_{200}$ by 
9 and 20 
per cent for WJ943.7 and 
11 and 29 
per cent for WJ943.5 
($r_{500}$ and $r_{200}$ themselves decrease
by 
3 and 7 
per cent for WJ943.7 and 
4 and 11
per cent for WJ943.5).
Given that the fractional errors on these masses would be
larger than for the isothermal case (because of errors on $\gamma$), the two 
scenarios produce masses consistent with each other within $1\sigma$.
Thus, within this statistical uncertainty the conclusions on $M_{grav}$ remain
unaffected by the presence of a temperature gradient.

\subsection{Masses of X-ray gas}\label{sec,gasmass}

To determine gas masses in the groups, the normalisation of the gas density 
profile must be established. For an isothermal ICM, this can be
directly derived from the best-fitting $\beta$-model for the surface 
brightness, using the spectral (in effect, luminosity) normalization in 
{\sc xspec},
\begin{equation}
A=10^{-14}(4\pi (D_A(1+z))^2 )^{-1} \int n_e n_H dV \mbox{ cm}^{-5},
\label{eq,xspec}
\end{equation}
where $D_A$ is the angular diameter distance to the source, and $n_e$ and
$n_H$ are the number densities of electrons and hydrogen, respectively.
With $n_e$ and $n_H$ following the best-fitting $\beta$--profile out to 
$r_{ext}$, their central values are found by evaluating the 
integral in eq.~(\ref{eq,xspec}), assuming  $n_e/n_H = 1.17$.
We find very similar central electron densities $n_{e,0}$ for the two groups 
of $3.15\pm 0.20 \times 10^{-3}$ $h^{1/2}$ cm$^{-3}$ (WJ943.7) and 
$2.91\pm 0.19 \times 10^{-3}$ $h^{1/2}$ cm$^{-3}$ (WJ943.5), with
errors obtained from the derived errors on the spectral normalization $A$.
When converting these number densities to mass densities 
$\rho = n_e \mu_e m_p = n_H \mu_H m_p$, $\mu_e = 1.17$ and $\mu_H = 1.40$
was assumed, appropriate for a fully ionized
$Z=0.3Z_{\odot}$ plasma (e.g.\ Mohr, Mathiesen \& Evrard 1999).
The total gas masses are finally found by simple volume integration of 
$\rho(r)$ inside $r_{ext}$.

In the non-isothermal case discussed above, $M_{gas}(r)$ changes because the 
derived gas density depends on emissivity and hence temperature. 
To quantify the magnitude of this change, we took for each group the 
best-fitting isothermal spectral models within $r_{ext}$ and changed 
the associated temperature within its extreme polytropic values at 
$T(r=0)$ and $T(r=r_{ext})$ (cf.\ Fig.~\ref{fig,tprof}), 
corrected to the 'true' value via eq.~(\ref{eq,ttrue}). 
This was done
subject to the constraint that the total 0.4--2.5 keV flux, from which
the surface brightness profile has been extracted, should remain constant.
The resulting change in spectral normalisation $A$ then translates, via 
eq.~(\ref{eq,xspec}),
into a maximum resulting error on the central gas density and hence total
gas mass. In all cases, these changes were found to be less than 5 per cent
within $r_{ext}$ relative to the isothermal case, i.e.\ lying within 
the $1\sigma$ statistical uncertainties from the spectral fits. 
We therefore assumed in 
the following that $M_{gas}$ remains unaffected by a change to the
polytropic models.

\begin{table*}
\caption{Derived gas masses, total gravitating masses, and gas mass fractions
within the derived extent of emission $r_{ext}$, $r_{500}$, and the 
approximate virial radius $r_{200}$. All values are given in the adopted 
$\Lambda$CDM cosmology of ($\Omega_m,\Omega_{\Lambda}$) = (0.3,0.7).}
\label{tab,mass}
\begin{tabular}{lccccccc}
\hline
Source & $r_{ext}/r_{200}$ & $r_{ext}/r_{500}$ & $M_{gas}(r_{ext})$ & $M_{tot}(r_{ext})$ &  $f_{gas}(r_{ext})$ & $f_{gas}(r_{500})$ &  $f_{gas}(r_{200})$ \\
 & & & ($h^{-5/2}M_{\odot}$) & ($h^{-1}M_{\odot}$) & ($h^{-3/2}$) & ($h^{-3/2}$) & ($h^{-3/2}$) \\ \hline
\vspace{.15cm}
WJ943.7 & 0.74 & 1.18 & $2.47^{+0.16}_{-0.16} \times 10^{12}$ & $3.82^{+0.36}_{-0.42}\times 10^{13}$ & $0.065^{+0.007}_{-0.008}$ & $0.058^{+0.007}_{-0.007}$  & $0.077^{+0.009}_{-0.010}$ \\ 
WJ943.5 & 0.66 & 1.07 & $5.94^{+0.37}_{-0.40} \times 10^{12}$ & $7.70^{+1.61}_{-1.32}\times 10^{13}$ & $0.077^{+0.017}_{-0.014}$ & $0.075^{+0.016}_{-0.014}$ & $0.089^{+0.019}_{-0.016}$ \\ \hline
\end{tabular}
\end{table*}

\section{Discussion}\label{sec,discus}

\subsection{Gas mass fractions}

The resulting gas mass fractions 
$f_{gas}=M_{gas}/M_{grav}$ within $r_{ext}$ and $r_{500}$ are also 
presented in Table~\ref{tab,mass}, 
along with the values resulting from extrapolating 
out to $r_{200}$. The associated errors were obtained from $M_{gas}$ and
$M_{grav}$ using standard error propagation.
For comparison with previous work on the topic, we note that the derived
gas mass fractions within ($r_{500}$,$r_{200})$ correspond to 
(0.049,0.065)$h^{-3/2}$ (WJ943.7) and (0.060,0.072)$h^{-3/2}$ (WJ943.5) 
for a flat standard CDM universe, 
and (0.053,0.070)$h^{-3/2}$ (WJ943.7) and (0.067,0.079)$h^{-3/2}$ 
(WJ943.5) for an open CDM universe with deceleration parameter $q_0 = 0$.
Based on a coherently analysed sample of 66 systems ranging
from individual galaxy haloes to massive clusters ($T=0.5-17$ keV), 
Sanderson et~al.\ (2003) derive a mean gas fraction within
$r_{200}$ of $<\!\!f_{gas}\!\!>=0.0785\pm0.006 h^{-3/2}$ (OCDM, $q_0 = 0$), 
with, in general, lower values for cool systems and a statistically 
significant trend of $f_{gas}$ increasing with $T$, a trend which is, 
however, subject to considerable scatter. 
The two groups studied here seem to conform to this trend, with  (in
this particular $q_0 = 0$ cosmology) WJ943.5 
lying close to this mean value and WJ943.7 having somewhat smaller 
$f_{gas}$.
In any given cosmology, the derived nominal gas mass fraction for the 
WJ943.5 group is also found 
to be comparable to mean values reported for more massive clusters 
($T>4$ keV) by Ettori \& Fabian (1999) and
Castillo-Morales \& Schindler (2003) and for intermediate-mass systems 
($T=2.5-5$ keV) by Mohr et al.~(1999) and Arnaud \& Evrard (1999). 
WJ943.7, however, displays a lower 
$f_{gas}$ than found in these studies, with a value very similar to, e.g., 
that found for the $T\simeq 2$ keV cluster A1983 (Pratt \& Arnaud 2003). 
For both groups, $f_{gas}$ is found to be 
clearly smaller than the values obtained for massive ($T >4$ keV) clusters 
(e.g., Mohr et al.~1999; Arnaud \& Evrard 1999; Grego et al.~2001; Sanderson 
et al.~2003). In summary, within $r_{200}$ WJ943.5 seems to be 
reasonably representative
of the general cluster population, whereas $f_{gas}$ of WJ943.7 is more
typical of values reported for slightly less massive systems situated at the 
interface between groups and clusters.

Using the polytropic models, we find that $f_{gas}$ is raised to 
0.061 (0.086) 
within $r_{500}$ ($r_{200}$) for WJ943.7, and 
0.079 (0.107) 
for WJ943.5. These values are larger than the nominal isothermal values by 
5
per cent within $r_{500}$ and between 
10 and 20
per cent within $r_{200}$. Even more so than for the total mass 
$M_{grav}$, they are comfortably consistent with the isothermal values 
within $1\sigma$. 
We stress that, barring any additional systematic errors, our
derived values within $r_{200}$ should be fairly reliable, since they involve
a relatively small degree of radial extrapolation compared to results for 
other 
groups. In this light it is interesting to note that our values are much 
closer to cluster values than often reported for small groups within smaller 
radii (e.g.\ David, Jones \& Forman 1995; Pedersen, Yoshii \& Sommer-Larsen
1997; Mohr et~al.\ 1999; Mushotzky et~al.\ 2003).

The modelled radial variation of $f_{gas}$ in both groups is shown in 
Fig.~\ref{fig,fgas}. 
For the polytropic models, $M_{grav}$ is lower at large 
radii, but $r_{500}$ and $r_{200}$, within which $M_{gas}$ is evaluated,
also decrease.
Since $f_{gas}$ clearly increases with $r$ even in the isothermal case, 
the relative impact of assuming a 
negative temperature gradient is therefore less pronounced at large radii for 
$f_{gas}$ than for $M_{grav}$
The gas mass fraction of both systems increases out to the radii
probed, with the standard implication that the intragroup gas is more 
extensively distributed than the dark matter. The consistency between the
isothermal and polytropic models strongly suggests that when taking the
errors into account, the overall radial behaviour of $f_{gas}$ shown in 
Fig.~\ref{fig,fgas} is a robust result.
The strong dependence of $f_{gas}$
 on radius suggested by the figure, 
particularly for the cooler WJ943.7, is furthermore corroborated by 
the results of \S \ref{sec,surfbright} for isothermal NFW fits to the surface 
brightness profiles, which, 
when scaled to $r_c$ or $r_{200}$, show NFW scale radii for the gas much 
larger than those of simulated and observed hotter systems
with a milder radial variation of $f_{gas}$ (cf.\ Ettori \& Fabian 1999).
The weak temperature dependence of the emissivity in the relevant temperature
and energy ranges (\S \ref{sec,noniso}) implies that, within the errors, 
this result holds equally well for the polytropic models.

\begin{figure}
\hspace{-0.5cm}
\epsfxsize=9cm
\epsfysize=7.5cm
\epsfbox{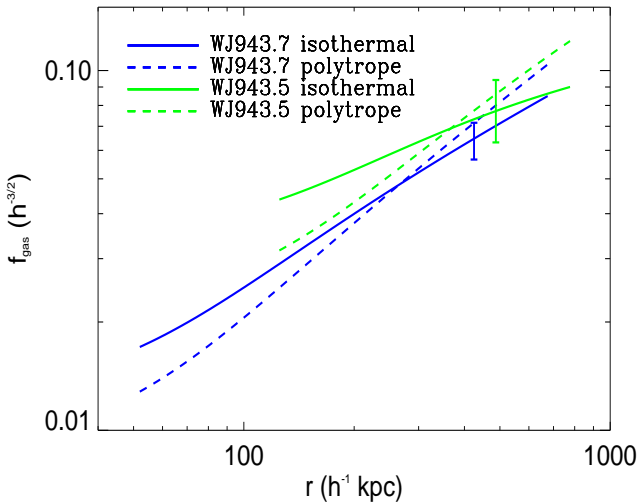}
\caption{Gas mass fraction $f_{gas}$ plotted from $r_c$ to $r_{200}$, using
two assumptions for the temperature profile. 
Dark curves for WJ943.7, light curves for WJ943.5. 
Solid lines represent isothermal gas, dashed lines the polytropic models 
discussed in the text. Typical error bars for the isothermal case are shown 
at $r=r_{ext}$, beyond which $f_{gas}$ is extrapolated.}
\label{fig,fgas}
\end{figure}



\subsection{Entropy}

The entropy of the ICM, here defined as $S\equiv T/n_e^{2/3}$, reflects the 
accretion history of the hot gas and as such is a prime 
indicator of the physical processes involved in the formation and evolution
of groups and clusters. 
Recent studies of the entropy in systems covering a wide range in masses 
have made it
clear that gravity-driven processes alone cannot account for the observed
entropy: Systems at all scales, most notably low-mass ones, show evidence
for excess entropy relative to that expected from purely gravitational 
heating and compression of the gas. 

Various mechanisms have been proposed to explain this excess entropy,
see e.g.\ the discussion in
Ponman, Sanderson \& Finoguenov (2003).  
We note here that the rather large values of
$f_{gas}$ found for our systems seem inconsistent with cooling-only
models, which require more than 50 per cent of the gas in low-$T$ systems 
like these to cool out of the diffuse hot phase 
(e.g.\ Dav\'e, Katz \& Weinberg 2002; Ponman et~al.\ 2003).

The radial variation of entropy cannot be computed in a 
model-independent way for the two groups studied here, given the poorly 
resolved temperature profiles. Using instead the 
best-fitting $\beta$-models along with the isothermal assumption, 
resulting central entropies are 
$78\pm 8$ and $118\pm 20$  $h^{-1/3}$ keV cm$^2$ for WJ943.7 and WJ943.5,
respectively, rising to 113 and 142 $h^{-1/3}$ keV cm$^2$ at a fiducial 
radius of $0.1r_{200}$. 

Self-similar scaling between systems of different masses would lead to
$S \propto (1+z)^{-2}T$. 
In Fig.~\ref{fig,entropy}, modelled profiles of $S(r)$ are plotted
for each group for both the isothermal and polytropic cases.
The entropy has been scaled by
$(1+z)^2/T$ so as to 'remove' the dependence on
mass and redshift in the self-similar case. 
In this representation the profiles of the two groups are seen to be fairly 
similar. 
The flattening in the entropy profiles
seen inside $0.1r_{200}$, due to the core in the gas density, is not a 
reliable result, since it is based on the assumption that $T(r)$ does not drop
within the central $\sim 50$~kpc. As can be seen from Fig.~\ref{fig,tprof}, 
our data do not allow us to resolve the temperature structure on such small 
scales. What is clear from our results, is that any isentropic core in these 
groups does not extend to $r\ga 0.2 r_{200}$, in contrast to the predictions 
of simple preheating models for such poor systems
(Tozzi \& Norman 2001, Babul et~al.\ 2002; Voit et~al.\ 2003).
This is consistent with recent results for other groups 
(Mushotzky et al.\ 2003; Pratt \& Arnaud 2003) and argues against significant
preheating in a simple spherical collapse scenario. 

Also plotted in Fig.~\ref{fig,entropy} (triangles) is the expectation for $S$
based on self-similar scaling of massive clusters (Ponman et~al.\ 2003).
For both groups, the scaled entropy exceeds that 
observed in massive clusters at $0.1r_{200}$. Further out our results are
more uncertain, but the modelled profiles suggest that this excess seems to 
persist 
out to at least $r_{500}$, where the scaled (isothermal) entropies of the
groups are a factor $\sim 2$ above the self-similar values. For our observed
gas density profiles, the polytropic indices of both groups would have to be
$\gamma \ga 1.40$ in order to avoid excess entropy at 
$r_{500}$. This is well beyond the values of $\gamma$ obtained for observed 
and simulated groups and clusters.
The derived entropy levels at $0.1r_{200}$ and $r_{500}$, and hence the 
associated excesses with respect to the massive clusters of 
Ponman et~al.\ (2003), are in good agreement with the best-fitting $S-T$ 
relations at the relevant radii of the {\em entire} group/cluster sample of 
Ponman et~al.\ (2003).



\begin{figure}
\begin{center}
\mbox{\hspace{-0.3cm}
\epsfxsize=9cm
\epsfysize=7.5cm
\epsfbox{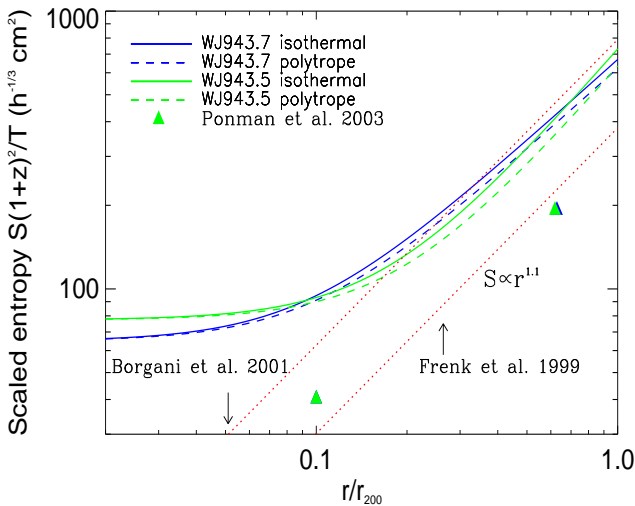}\hspace{0cm}}
\end{center}
 \caption{
Scaled entropy $S(1+z)^2/T$ as a function of $r/r_{200}$, using two 
assumptions for the temperature profile.
Triangles mark the expected values at $0.1 r_{200}$ and
$r_{500}$ based on a self-similar scaling from the values observed in hot
clusters.
Sloping dotted lines represent the self-similar profile expected from
shock heating in a spherical smooth accretion scenario, $S\propto r^{1.1}$, 
normalised at $r_{200}$ to the preheated $T\approx 2$ keV cluster simulation 
of Borgani et~al.\ (2001),
and to the simulated $T\approx 6.5$ keV 
cluster of Frenk et~al.\ (1999)
which includes no heating or cooling effects.}
\label{fig,entropy}
\end{figure}

Spherical shock heating models and most cosmological simulations
produce entropy profiles with a power law form $S\propto r^{1.1}$ 
(e.g. Tozzi \& Norman 2001). 
Two lines of this slope have also been
plotted in Fig.~\ref{fig,entropy}. The lowest line, which agrees fairly
well with the triangles representing observations of rich clusters, is
normalised to the scaled mean value at $r_{200}$ of the 
$M \simeq 5\times 10^{14} h^{-1} M_{\odot}$ simulated 'Santa Barbara' 
cluster (Frenk et~al.\ 1999), which included only gravitational physics.
The profile has been scaled using an estimated 
emission-weighted temperature of $T_{ew}=6.5$ keV, derived from its 
mass-weighted value of $T_{mw}=4.7$ keV assuming a ratio 
$T_{ew}/T_{em}\approx 1.4$, as suggested by results from simulations 
(Borgani et~al.\ 2002) and observations (Sanderson et~al.\ 2003).
While only a suggestive result, this profile clearly fails to reproduce our
modelled entropy levels at large radii.
Also shown is an $S\propto r^{1.1}$ line normalised at $r_{200}$ to the 
result obtained for a $T=2.1$ keV cluster in numerical simulations of cluster 
formation by Borgani et~al.\ (2001). These simulations include preheating at a
level of $45 h^{-1/3}$ keV cm$^2$,
and do reproduce the large
entropy excesses hinted at for our groups at $r_{500}$ without forming a
distinct isentropic core.

Whereas the absence of a large isentropic core is in direct conflict with 
simple smooth accretion of preheated gas, the 
excess entropy at large radii 
suggested by our analytical profiles
may, in fact, be explained by such models
(for details, see Voit et~al.\ 2003 and Voit \& Ponman 2003).

\section{Conclusions}\label{sec,concl}

The main results of this work may be summarized as follows.

\begin{itemize}
\item Using sensitive {\it XMM-Newton} observations, we have mapped the X-ray 
emission out to, on average, 70 per cent of the estimated virial radii 
$r_{200}$ in 
two X-ray selected galaxy groups at redshifts $z=0.18$ and $z=0.256$.
We find the X-ray surface brightness profiles of both systems to be well 
described by standard $\beta$-models out to the radii probed, with indices of
$\beta=0.49$ and 0.63. Fits to isothermal NFW profiles for the gas density
indicate that the surface brightness data do not rule out a steepening in the
density profile at large radii as might be expected on physical grounds.
However, adequate fits for such models can only be obtained with a 
very large scale radius, and
models in which gas traces mass are strongly ruled out for both groups.

\item Emission-weighted mean temperatures, determined within the same radii 
as the surface brightness profiles, are found to be $kT=1.7\pm0.1$ and 
$2.4\pm0.4$ keV, with corresponding gas
metallicities roughly 0.3 solar. The temperature profiles are poorly 
resolved but are consistent with isothermality within 90 per cent confidence.

\item Radial profiles of gas mass, gravitating mass, and gas mass
fraction have been determined assuming both isothermal and polytropic gas
distributions allowed by the data. In both cases, the gas mass fraction 
clearly 
increases with radius, implying a more extended gas distribution than that of
the total mass. Within the derived extent of emission, gas mass
fractions are $0.065h^{-3/2}$ and $0.077h^{-3/2}$ in the two groups, 
considerably higher than values reported in the literature for other groups 
within smaller radii. Extrapolated gas mass fractions at $r_{200}$ 
of $0.077h^{-3/2}$ and $0.089h^{-3/2}$ are slightly lower, but comparable to,
values found for more massive clusters. 
These results emphasize the importance of extending group studies to large 
radii, in order to obtain reliable information on the amount and distribution
of baryonic and dark matter in these systems.


\item The radial variation of gas entropy in the two groups shows no evidence
for a large isentropic core, at variance with predictions from simple 
preheating scenarios. The poorly resolved temperature profiles prohibit any
firm conclusions on the entropy behaviour at large radii, but evidence for 
excess entropy with respect to that expected from 
self-similar scaling of observed massive clusters is seen out to at least 
$r_{500}$. Such a large-radius excess would be
consistent with the predictions of some simulations including heat sources,
and with smooth accretion 
theory. As an alternative for the origin of excess entropy at all scales,
radiative cooling alone is ruled out by 
the relatively high gas mass fractions derived for these systems.

\end{itemize}

\section*{Acknowledgments}
We thank Andy Read and Ben Maughan for many useful discussions and use of 
their X-ray analysis scripts, Laurence Jones for discussions of the WARPS 
data, and Gary Wegner (Dartmouth College) and Harald Ebeling 
(Univ.\ of Hawaii) for letting us use these data prior to publication. 
Tadayuki Kodama is thanked for providing us 
with his elliptical galaxy evolution models.
This work acknowledges the use of the Starlink facilities at Birmingham,
the Lyon and NASA Extragalactic Databases, and the ROSAT All-Sky
Survey and 2-Micron All-Sky Survey databases. 
JR acknowledges support by the Danish Natural Science Research Council (SNF).


\label{lastpage}

\end{document}